\documentclass[12pt]{article}
\usepackage{fullpage, amsmath}
\usepackage{graphicx}
\usepackage{enumerate}
\usepackage{natbib}
\usepackage{url} 
\usepackage{amssymb}
\usepackage{color, multirow, lscape}
\usepackage{xr}

\newcommand{\real}{\mathbb R}

\newcommand{\E}{\mathbb{E}}

\newcommand{\Cov}{\textnormal{Cov}}
\newcommand{\tr}{\mathrm{tr}}

\def\bB{\mathbf{B}}
\def\bb{\mathbf{b}}

\def\bD{\mathbf{D}}

\def\bI{\mathbf{I}}

\def\bS{\mathbf{S}}
\def\bG{\mathbf{G}}

\def\bx{\mathbf{x}}

\def\by{\mathbf{y}}

\def\bt{\mathbf{t}}
\def\bs{\mathbf{s}}

\def\bz{\mathbf{z}}

\def\balpha{\boldsymbol{\alpha}}
\def\bmu{\boldsymbol{\mu}}
\def\bxi{\boldsymbol{\xi}}
\def\bbeta{\boldsymbol{\beta}}

\def\bbbeta{\boldsymbol{\eta}}
\def\bepsilon{\boldsymbol{\epsilon}}

\def\bOmega{\boldsymbol{\Omega}}

\def\bsigma{\boldsymbol{\sigma}}
\def\bgamma{\boldsymbol{\gamma}}
\def\bSigma{\boldsymbol{\Sigma}}
\def\bPhi{\boldsymbol{\Phi}}
\def\bPsi{\boldsymbol{\Psi}}

\def\bzeta{\boldsymbol{\zeta}}
\def\bTheta{\boldsymbol{\Theta}}
\def\btheta{\boldsymbol{\theta}}
\def\beeta{\boldsymbol{\eta}}

\def\N{\mathcal{N}}
\def\F{\mathcal{F}}
\def\C{\mathcal{C}}

\title{Joint Model for Survival and Multivariate Sparse Functional Data with Application to a Study of Alzheimer's Disease}

\author{Cai Li$^{1,*}$, Luo Xiao$^{2}$, and Sheng Luo$^{3}$ \\
$^{1}$Department of Biostatistics, Yale University\\
$^{2}$Department of Statistics, North Carolina State University\\
$^{3}$Department of Biostatistics and Bioinformatics, Duke University\\
$^{*}$cai.li@yale.edu}

\begin{document}
\maketitle

\abstract
Studies of Alzheimer's disease (AD) often collect multiple longitudinal clinical outcomes, which are correlated and predictive of AD progression. It is of great scientific interest to investigate the association between the outcomes and time to AD onset. 
We model the multiple longitudinal outcomes as multivariate sparse functional data and propose a functional joint model linking multivariate functional data to  event time data. In particular, we propose a  multivariate functional mixed model (MFMM) to identify the shared progression pattern and outcome-specific progression patterns of the outcomes, which enables more interpretable modeling of  associations between outcomes and AD onset.
The proposed method is applied to the Alzheimer's Disease Neuroimaging Initiative study (ADNI) and the functional joint model sheds new light on inference of five longitudinal outcomes and their associations with AD onset. Simulation studies also confirm the validity of the proposed model.


\section{Introduction}
\label{sec:intro}
Alzheimer's disease (AD) is the most prevalent neurodegenerative disorder, can often be characterized by accelerated metal degradation over time,  and may ultimately progress to dementia. 
In the year of 2017, AD was the sixth leading cause of death in the United States with 121,494 recorded deaths \citep{alzheimer20192019}.
Great efforts have been dedicated to advancing early detection of AD.

The motivating data are from the Alzheimer's Disease Neuroimaging Initiative (ADNI) with the primary goal of investigating whether multimodal data can be combined to measure the progression of AD \citep{weiner2017recent}, and are publicly available at \url{http://adni.loni.ucla.edu}. 
We are interested in jointly modeling clinical variables, multiple longitudinal outcomes measured intermittently and time to AD onset or drop-out. Throughout the paper, AD onset refers to a clinical declaration of probable AD based on cognitive symptoms.
We consider five longitudinal biomarkers commonly measured in AD studies. 
Among the five biomarkers, 
high values of Disease Assessment Scale-Cognitive 13 items (ADAS-Cog 13) and Functional Assessment Questionnaire (FAQ) reflect severe cognitive decline, whereas low values of Rey Auditory Verbal Learning Test immediate recall (RAVLT-immediate), Rey Auditory Verbal Learning Test learning curve (RAVLT-learn) and Mini-Mental State Examination (MMSE) indicate a high risk for developing AD.

\begin{figure}[htp]
	\centering	
	\scalebox{0.35}{
		\includegraphics[]{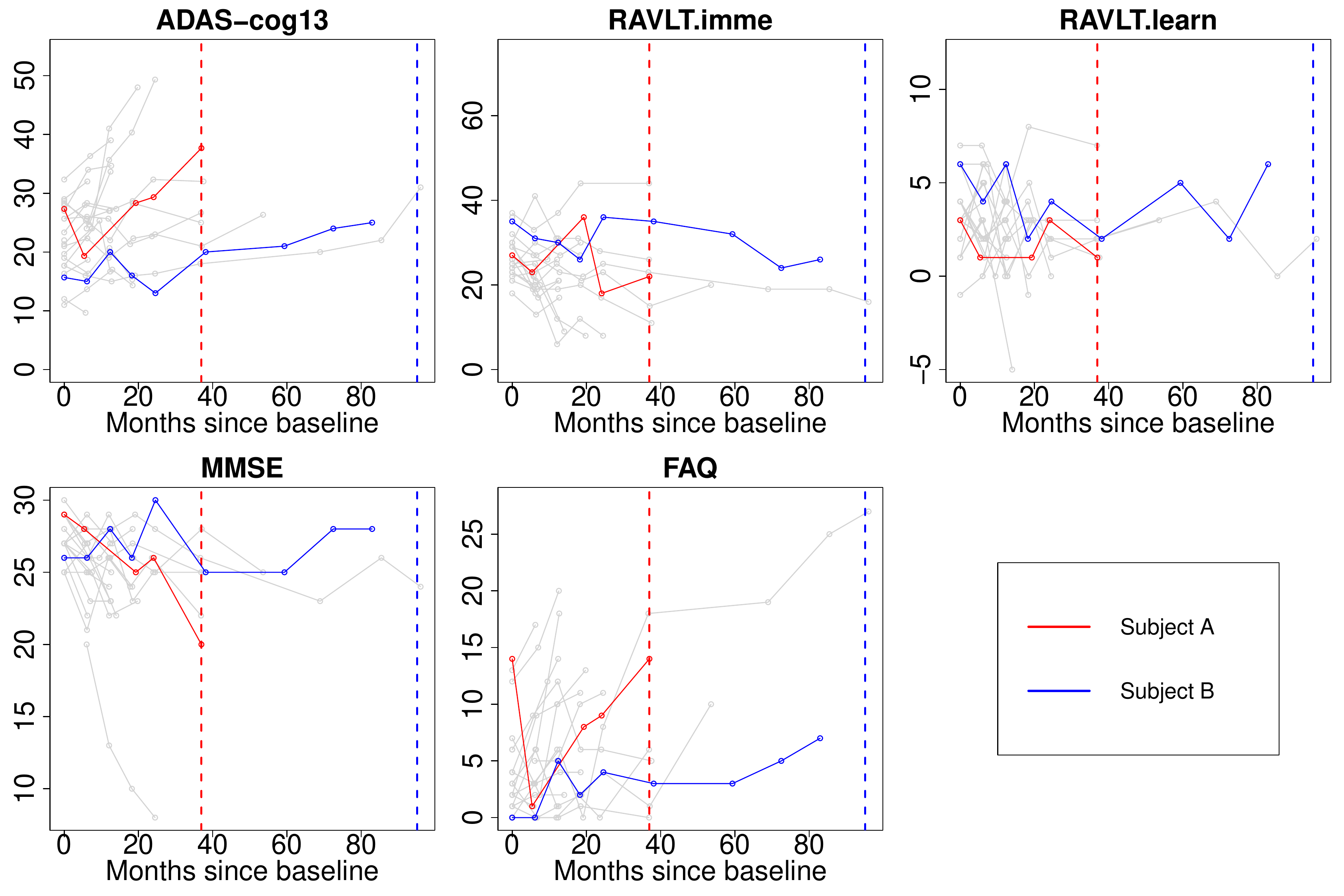}
	}
	\caption{\label{fig:long}Five observed longitudinal biomarkers of two subjects. Vertical lines: time to AD onset; gray lines: longitudinal outcomes for 20 other subjects.}
\end{figure}

Figure \ref{fig:long} presents spaghetti plots of the five longitudinal biomarkers and highlights profiles and time to AD onset for two subjects. Subject A has a more acute cognitive decline compared to Subject B. Indeed, subject A has faster increasing values of ADAS-Cog 13 and FAQ, faster decreasing values of RAVLT-immediate, RAVLT-learn, and MMSE as well as an earlier AD onset around month 40.
The plots motivate our research: (a) subjects with more severe cognitive impairment seem more likely to progress to AD; and (b) the multiple longitudinal outcomes may be correlated. 
In addition, the observations do not always show clear linear trends and the trends across outcomes may not be synchronized, suggesting of potentially heterogeneous patterns.

Parametric models are often used in the joint modeling literature \citep{tsiatis2004joint}.
\cite{de1994modelling} introduced a shared random effects model, which links longitudinal data to event time data via a set of time-invariant random effects.
\cite{wulfsohn1997joint} proposed a shared latent process model, which models the instantaneous effect of longitudinal data to event time data.
Extensions of the univariate joint models to multiple longitudinal outcomes include \cite{henderson2000joint} and \cite{lin2002maximum}. 
However, limitations exist: (1) the parametric models are incapable of modeling complex nonlinear trends of longitudinal outcomes; (2) assuming a specific structure of the correlation may be subject to model misspecification. 
Recently, joint modeling of event time data and functional data has drawn some attention.
\cite{yao2007functional} proposed a shared latent process model. 
\cite{yan2017dynamic} proposed a shared random effect model with a two-step estimation, 
and \cite{ye2015joint} proposed a model for  baseline longitudinal patterns and interval-censored event time data.
To clarify,  joint modeling here refers to the situation where the domain of function is longitudinal time. By contrast, Cox regression models  where functional data are used as baseline covariates have been extensively studied; see \cite{kong2018flcrm} and references therein.
Nevertheless, none of the above works considered joint modeling of multivariate functional data and event time data.

To capture the heterogeneity of patterns in the outcomes as well as correlations among them, a popular method is  multivariate functional principal component analysis \citep{happ2018multivariate}.  Then  a Cox regression model may be adopted to link the outcomes and AD onset via the functional principal component socres.
However, there exist serious computational issues
for joint model estimation. Multivariate functional principal component analysis (MFPCA) models the mean functions, auto-covariance functions for within-function correlations, and cross-covariances for between-function correlations nonparametrically. Thus, for $J$-dimensional functional data, there are $J$ univariate functions and $J^2$ bivariate functions to estimate.
If we use tensor product splines to approximate the bivariate fucntions and each marginal basis is of dimension $c$, then it leads to $J^2 c^2$ parameters to estimate,
which is computationally prohibitive for joint model estimation if $J$ is more than 2 and infeasible for large $J$.
Moreover, MFPCA is mostly used for dimension reduction and the resulting multivariate eigenfunctions and associated scores are often difficult to interpret. 
In particular, MFPCA does not explicitly model the correlation between outcomes as is often done in parametric models for multivariate longitudinal data \citep{Verbeke2014}.

We propose a new multivariate functional mixed model (MFMM) for multivariate functional data and subsequently a new functional joint model for linking multivariate functional data and event time data.
The advantages of the proposed methods include: 
(1) the MFMM  retains the flexibility of functional data methods for capturing nonlinear patterns in longitudinal outcomes and models the correlation between outcomes via a shared latent process; 
(2) compared to $J^2$ bivariate covariance functions for MFPCA,  MFMM requires only two bivariate covariance functions, which makes joint model estimation feasible; 
(3) MFMM explicitly separates the shared latent process, common to all outcomes, from the outcome-specific latent processes, unique to each outcome, and thus greatly enhances model interpretability; 
(4) MFMM enables a flexible Cox regression model which not only evaluates the effects of the shared latent process to survival risk but also identifies the additional contribution of each outcome to survival risk.

The remainder of the paper is organized as follows.
Section \ref{sec:models} introduces the proposed multivariate functional mixed model as well as the joint model for disease progression and survival. 
Section \ref{sec:estimation} describes a two-step estimation method and the proposed joint estimation method.
Section \ref{sec:mod_sel} presents model selection for the proposed model.
Section \ref{sec:application} applies the proposed model to the ADNI data.
Section \ref{sec:sim} examines the numerical properties of the proposed method through simulations.
Section \ref{sec:discussion} concludes this work with some discussion.
Technical details and extra results for numerical studies are included in the Web Appendices.

\section{Model}
\label{sec:models}
Let $Y_{ijk}$ be the $k$th observation of the $j$th outcome (biomarker) measured intermittently at time $t_{ijk}$ for subject $i$
with $1\leq k\leq m_{ij}$, $1\leq j\leq J$, and $1\leq i\leq n$. 
So there are $n$ subjects and subject $i$ has $m_{ij}$ observations for the $j$th outcome.
In our data application, we model the five longitudinal outcomes ADAS-Cog 13, RAVLT-immediate, RAVLT-learn, MMSE, and FAQ (plotted in Figure \ref{fig:long}) as $Y_{ijk}$ with $j = 1, \cdots, 5$, respectively. 
The survival time for the $i$th subject is denoted by $S_i$ and is  assumed to be  subject to independent right censoring with censoring time denoted by $C_i$. Let $T_i = \min(S_i, C_i)$ and $\Delta_i = 1_{\{S_i \leq C_i\}}$.
Note that $t_{ijk} \in [0,T_i]$, meaning no observations  after $T_i$. Assume $T_i \leq \tau$ for all $i$, where $\tau$ is the length of study follow-up.
Denote by $\bz_i = (Z_{i1},\ldots, Z_{iP})\in\real^P$ the vector of baseline covariates.

\subsection{Multivariate Functional Mixed Model}
We propose a multivariate functional mixed model (MFMM) for the multivariate latent process $\{X_{i1}(t),\ldots, X_{iJ}(t)\}$.
The MFMM is of the form
\begin{equation}
\label{eq:model_long}
Y_{ijk} = X_{ij} (t_{ijk}) + \epsilon_{ijk},\quad
X_{ij}(t) = \mu_j(t) + \beta_j \left\{U_{i}(t) + W_{ij}(t)\right\}, 
\end{equation}
where $\epsilon_{ijk}$ are  random noises
so that the longitudinal outcome $Y_{ijk}$ is a proxy observation of the true latent stochastic process $X_{ij}(t)$ evaluated at time $t_{ijk}$.
The smooth latent  process $X_{ij}(t)$ is decomposed into three components.
First, $\mu_j(t)$ is the fixed mean function for outcome $j$. For simplicity, we assume that the mean function only depends on the longitudinal time but it may depend on the baseline covariates, which can be incorporated easily using, for example, additive models.
The continuous latent
profile $U_i(t)$, common to multiple outcomes, is a subject-specific random deviation from the  mean functions. $U_i(t)$ captures the subject-specific disease progression pattern and correlation among outcomes. It represents
subject $i$'s unique latent disease status at time $t$ manifested by multiple outcomes and can be specified so that  a higher value indicates more
severe status. The outcome-specific scaling parameter $\beta_j$ is the expected increase in outcome $j$ for one unit increase
in $U_i(t)$. If two outcomes are negatively correlated, their $\beta_j$s have different signs. 
$W_{ij}(t)$ is the subject- and outcome-specific random deviation from the outcome-specific mean and it characterizes subject $i$'s outcome-specific progression pattern.
By multiplying the scaling
parameters $\beta_j$, $W_{ij}(t)$ are comparable across outcomes. 

We model $U_i(\cdot)$ and $W_{ij}(\cdot)$ via two zero-mean Gaussian processes with covariance functions $\C_0(s,t) = \Cov\{U_i(s), U_i(t)\}$ and $\C_1(s,t) = \Cov\{W_{ij}(s), W_{ij}(t)\}$, respectively. Consider  the spectral decomposition of the covariance functions,
$\C_0(s,t) = \sum_{\ell} d_{0 \ell} \phi_{\ell}(s) \phi_{\ell}(t)$ and $\C_1(s,t) = \sum_{\ell} d_{1 \ell} \psi_{\ell}(s) \psi_{\ell}(t)$,
where $d_{0 1} \geq d_{0 2} \geq \cdots $ and $d_{1 1} \geq d_{1 2} \geq \cdots $ are the ordered eigenvalues, and $\phi_{\ell}(\cdot)$ and $\psi_{\ell}(\cdot)$ are the associated orthonormal eigenfunctions satisfying $\int_0^{\tau} \phi_{\ell}(t)\phi_{\ell^{\prime}}(t) dt = \int_0^{\tau} \psi_{\ell}(t)\psi_{\ell^{\prime}}(t) dt = 1_{\left\{\ell=\ell^{\prime}\right\}}$.
Then the Karhunen-Lo\`eve representations of $U_i(t)$ and $W_{ij}(t)$ are
$U_i(t) = \sum_{\ell \geq 1} \phi_{\ell}(t)\xi_{i \ell}$, $W_{ij}(t) = \sum_{\ell \geq 1} \psi_{\ell}(t)\zeta_{ij \ell}$,
where  $\xi_{i \ell} \sim \N(0, d_{0 \ell})$ are eigen scores and independent over $\ell$, and $\zeta_{ij \ell} \sim \N(0, d_{1 \ell})$ are defined similarly and independent over $j$ and $\ell$. 
The eigenfunctions $\phi_{\ell}(t)$ and $\psi_{\ell}(t)$ represent the changing patterns of the latent disease profiles
and the random scores $\xi_{i \ell}$ and $\zeta_{ij \ell}$ determine how strongly subject $i$'s latent disease profile follows those  patterns.
In practice, we assume there are only a finite number of  patterns so that 
$
U_i(t) = \sum_{\ell = 1}^{L_0} \phi_{\ell}(t)\xi_{i \ell}, 
W_{ij}(t) = \sum_{\ell = 1}^{L_1} \psi_{\ell}(t)\zeta_{ij \ell},
$
where $L_0$ and $L_1$ are finite numbers.
We shall treat $L_0$ and $L_1$ as tuning parameters and select them through data adaptive methods; see  Section \ref{sec:mod_sel} for details.
We assume that the random noises $\epsilon_{ijk}$ are independent and normally distributed with zero mean and variance $\sigma_{j}^2$. Finally, $U_i(t)$, $W_{ij}(t)$ and $\epsilon_{ijk}$ are assumed independent between subjects and across each other. 
\begin{figure}[htp]
	\centering	
	\scalebox{0.35}{
		\includegraphics[]{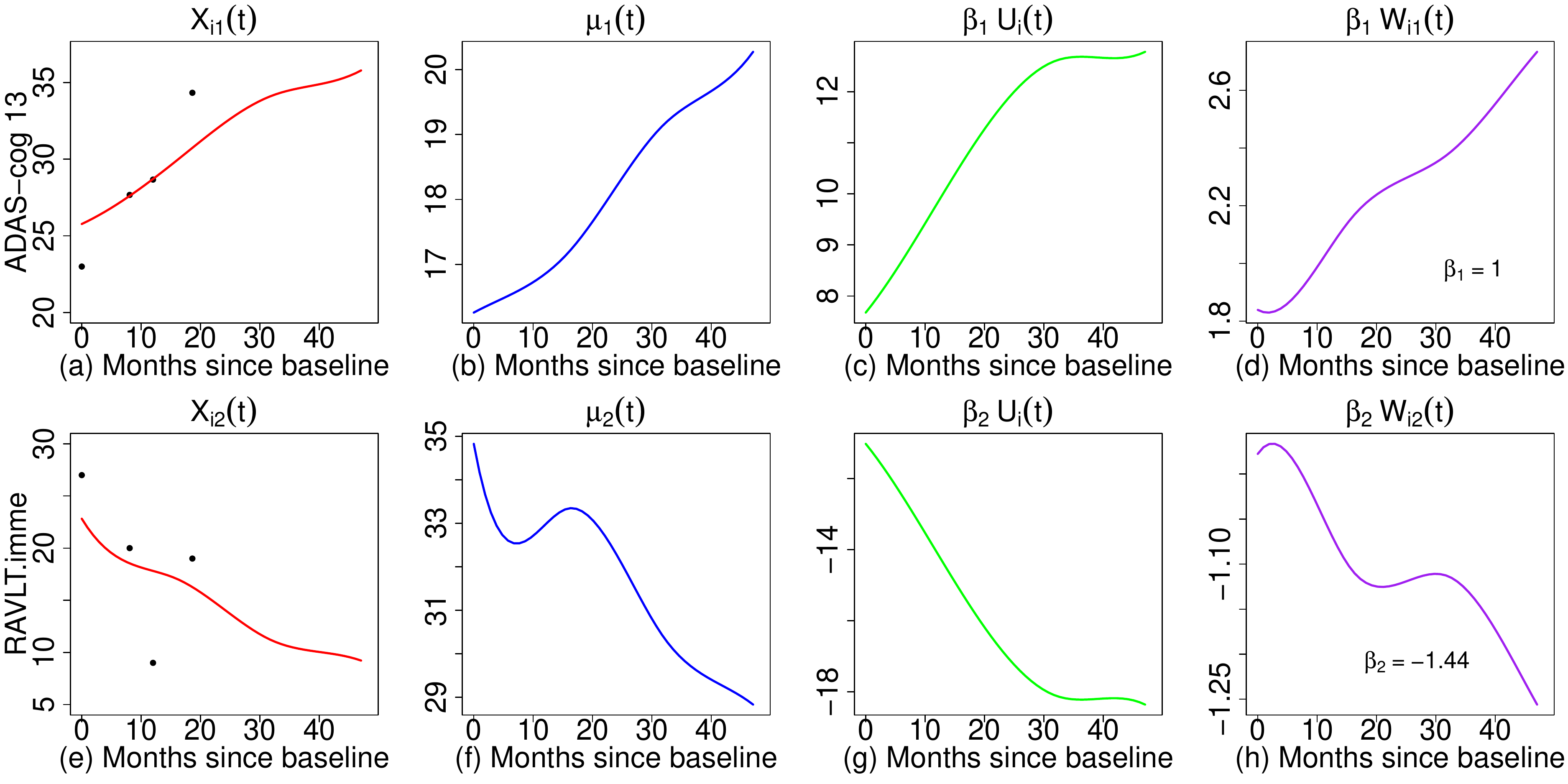}
	}
	\caption{\label{fig:MMFPCA}Estimate of each component in model \eqref{eq:model_long} for outcomes
		ADAS-Cog 13 (a-d) and RAVLT-immediate (e-h) for one subject in the ADNI study. 
		(a) and (e): observed ADAS-Cog 13 and
		RAVLT-immediate values (black dots) and the latent processes
		$X_{ij}(t)$; (b) and (f):  mean functions $\mu_j(t)$; (c)
		and (g): shared latent disease profile $U_i(t)$ multiplied by $\beta_j$ ; (d) and (h):
		outcome-specific deviations $W_{ij}(t)$ multiplied by $\beta_j$.}
\end{figure}

We illustrate the proposed MFMM by fitting five biomarkers in the ADNI study and present two outcomes ADAS-Cog 13 and RAVLT-immediate in Figure \ref{fig:MMFPCA}; see Section \ref{sec:application} for model fitting details.
In Figure \ref{fig:MMFPCA},
the estimate or prediction of each component in model \eqref{eq:model_long} is visualized for the two outcomes of one subject. To make model identifiable, we set $\beta_1$ = 1 for the outcome ADAS-Cog 13. 
Figures \ref{fig:MMFPCA}(a) and \ref{fig:MMFPCA}(e) present the fitted latent processes by MFMM for both outcomes. The subject shows steady worsening in cognitive function $X_{i1}(t)$ (increasing ADAD-Cog 13 in Figure \ref{fig:MMFPCA}(a)), which can be decomposed into increasing mean cognitive function $\mu_1(t)$ (Figure \ref{fig:MMFPCA}(b)),  deteriorating (increasing) latent disease profile $U_i(t)$ (Figure \ref{fig:MMFPCA}(c)), and positive outcome-specific progression $W_{i1}(t)$ (Figure \ref{fig:MMFPCA}(d)). Similar interpretation can be made to the outcome RAVLT-immediate and decreasing patterns indicate AD progression.

Model \eqref{eq:model_long} allows us to explicitly model the shared latent disease profile $U_i(t)$ between the outcomes and outcome-specific profile $W_{ij}(t)$. While it has a similar multilevel decomposition structure as in multilevel FPCA \citep{di2009multilevel}, there exist significant differences. The proposed MFMM accommodates  outcome heterogeneity: (1) The scaling parameters reduce heterogeneity of the multiple functions, which may measure quantitatively very different features of subjects. For example, $\beta_2$ for RAVLT-immediate is estimated as a negative number and it changes the direction of $U_i(t)$ so that it is negatively correlated with ADAS-Cog 13; and (2) the outcome-specific progression further accommodates data heterogeneity, such as, $\beta_1 W_{i1}(t)$ shows larger deviation from zero toward AD onset as compared with $\beta_2 W_{i2}(t)$, suggesting more severe disease progression in ADAS-Cog 13 than in RAVLT-immediate of the subject. 
Compared with multivariate FPCA, MFMM borrows its idea but further accounts for outcome-specific patterns,
which leads to theoretical and practical advantages: (1) MFMM gives a more interpretable model of multiple outcomes by separating the shared component, which models the correlation between outcomes, from outcome-specific components, which model the patterns of outcomes that are uncorrelated from other outcomes. 
By contrast, MFPCA only considers the shared component by reducing the multiple outcomes into a set of uncorrelated scores; 
and (2) by imposing a parsimonious model, the number of auto- and cross-covariance functions to be estimated are not increasing with the number of outcomes. 
This is a reasonable compromise between computability and medical fidelity,
which greatly alleviates computational burden and thus makes the joint estimation feasible.
In addition, MFMM may be regarded as a nonparametric extension of parametric multi-level decomposition model for multivariate longitudinal data \citep{Verbeke2014}.

We derive from \eqref{eq:model_long} that
\begin{equation}
\label{eq:moment:generic}
\C_{jj'}(s,t):= \Cov\{X_{ij}(s), X_{ij'}(t)\}=
\beta_j \beta_{j'} \C_0(s, t) + \beta_j^2 1_{\{j=j'\}} \C_1(s, t).
\end{equation}
If $j\neq j'$, 
$\C_{jj'}(s,t) = \beta_j \beta_{j'} \C_0(s,t)$. For model identifiability, we let $\beta_1 = 1$. Then it can be shown that for $J\geq 2$, $\beta_j$ and $\C_0(\cdot, \cdot)$
can be uniquely determined by \eqref{eq:moment:generic} using the equations with $j\neq j'$. 
See Web Appendix A for proofs and Web Appendix B for the covariance structure relating MFMM to MFPCA.

\subsection{Joint Model for Disease Progression and Survival}
To model the survival time, we use the proportional hazards model
\begin{eqnarray}
h_i(t) = h_0(t) \exp\{\bz_i^{\top} \bgamma_z + \F(\bx_i, t)\}, \label{eq:model_cox}
\end{eqnarray}
where $h_0(\cdot)$ is the baseline hazard function, $\bgamma_z$ is the coefficient vector corresponding to baseline covariates $\bz_i$, $\bx_i$ is the collection of all outcomes for subject $i$, and $\F(\bx_i, t)$ is the regression term of multiple latent processes at time $t$.
We consider the framework of shared random effects models \citep{wu1988estimation,de1994modelling} as it takes into account the entire history of the latent processes and let $\F(\bx_i, t) = \sum_{\ell=1}^{L_0}\xi_{i\ell}\gamma_{0 \ell} + \sum_{j=1}^{J}\sum_{\ell=1}^{L_1}\zeta_{ij\ell}\gamma_{1 j \ell}$, where $\gamma_{0 \ell}$ and $\gamma_{1 j \ell}$ are the coefficients corresponding to the shared and outcome-specific latent profiles, respectively. The hazard model extends the model in  \cite{yan2017dynamic} for univariate functional data to multiviarate functional data. 

\subsection{Likelihood of Joint Model}\label{sec:lik}
For model estimation, we now derive the likelihood function of the multivariate longitudinal outcomes and the event time data. 

We shall introduce some notation, which will be used throughout the rest of the paper.
Let $\bxi_i = (\xi_{i1}, \cdots, \xi_{iL_0})^{\top}$ be the vector of  scores for the shared latent profile $U_i(t)$. Then  $\bxi_i \sim \N(\mathbf{0}, \bD_0)$, where $\bD_0 \in \real^{L_0 \times L_0}$ is a diagonal matrix with $d_{0 \ell}$ the $\ell$th diagonal element.
Let $\bzeta_{ij} = (\zeta_{ij1}, \cdots, \zeta_{ijL_1})^{\top}$ be the vector of  scores for the outcome-specific latent profiles $W_{ij}(t)$.
Then
$\bzeta_{ij}\sim \N(\mathbf{0}, \bD_1)$, where $\bD_1 \in \real^{L_1 \times L_1}$ is a diagonal matrix with $d_{1 \ell}$ the $\ell$th diagonal element. 
Similarly, let $\by_{ij} = \left(Y_{ij1}, \cdots, Y_{ij m_i}\right)^{\top}$ be the vector of observations for the $j$th outcome and $\by_i = \left(\by_{i1}^{\top}, \cdots, \by_{i J}^{\top}\right)^{\top}$. 
Let $\bmu_{ij} = \left\{\mu_{ij}(t_{i1}), \cdots, \mu_{ij}(t_{i m_i}) \right\}^{\top}$ be the vector of the $j$th mean function at the observed time points. 
Let $\bPhi(t) = \left\{\phi_1(t), \cdots, \phi_{L_0}(t) \right\}^{\top}$ and $\bPsi(t) = \left\{\psi_1(t), \cdots, \psi_{L_1}(t) \right\}^{\top}$.
Then let $\bPhi_i = \left\{\bPhi(t_{i1}), \cdots, \bPhi(t_{im_i}) \right\}^{\top}$ and $\bPsi_i = \left\{\bPsi(t_{i1}), \cdots, \bPsi(t_{im_i}) \right\}^{\top}$ be the matrices of eigenfunctions evaluated at the observed time points.
Denote by $\bx_{ij}  = \left\{X_{ij}(t_{i1}), \cdots, X_{ij}(t_{i m_i})\right\}^{\top}$ 
the vector of the $j$th outcome evaluated at the observed time points without measurement errors, note that $\bx_{ij} = \beta_j (\bPhi_i \bxi_i + \bPsi_i \bzeta_{ij})$, and $\bx_i = \left(\bx_{i1}^{\top}, \cdots, \bx_{i J}^{\top}\right)^{\top}$. Finally, let $\bt_i = (t_{i1}, \cdots, t_{i m_i})^{\top}$ be the vector of the observed time points, and  $\bSigma_i = \text{blockdiag}(\sigma_1^2\bI_{m_i}, \cdots, \sigma_J^2\bI_{m_i})$. 

First,
the conditional likelihood of multivariate longitudinal data is
\begin{eqnarray}
\label{eq:llong}
f(\by_i | \bx_i, \bt_i, \bSigma_i) &=& ( |2 \pi \bSigma_i|)^{-\frac{1}{2}} \exp\left\{-\frac{1}{2}(\by_i - \bx_i)^{\top}\bSigma_i^{-1}(\by_i - \bx_i)\right\}
\end{eqnarray}
and $f(\bxi_i|\bD_0) = ( |2 \pi \bD_0|)^{-\frac{1}{2}} \exp\left(-\frac{1}{2}\bxi_i^{\top}\bD_0^{-1}\bxi_i\right)$, 
$f(\bzeta_{ij}|\bD_1) = ( |2 \pi \bD_1|)^{-\frac{1}{2}} \exp\left(-\frac{1}{2}\bzeta_{ij}^{\top}\bD_1^{-1}\bzeta_{ij}\right)$.
Next, the  conditional likelihood of time-to-event data is given by
\begin{eqnarray}
\label{eq:lcox}
f(T_i, \Delta_i |h_0, \bz_i, \bx_i, \bgamma_z, \bgamma_\eta) = \left\{h_0(T_i) \exp(\bz_i^{\top}\bgamma_z + \bbbeta_i^{\top} \bgamma_\eta)\right\}^{\Delta_i} \exp\left\{-\int_{0}^{T_i} h_0(u) \exp(\bz_i^{\top}\bgamma_z +\bbbeta_i^{\top}\bgamma_\eta ) du\right\},
\end{eqnarray}
where  $\bbbeta_i = \left(\bxi_i^{\top}, \bzeta_{i1}^{\top}, \cdots, \bzeta_{iJ}^{\top}\right)^{\top}$, $\bgamma_\eta = \left( \bgamma_0^{\top}, \bgamma_{11}^{\top}, \cdots, \bgamma_{1J}^{\top}\right)^{\top}$, $\bgamma_0 = (\gamma_{0 1}, \cdots, \gamma_{0 L_0})^{\top}$, and  $\bgamma_{1 j} = (\gamma_{1 j 1}, \cdots, \gamma_{1 j L_1})^{\top}$ for all $j$.
As  the multivariate longitudinal data and the time-to-event data are conditionally independent given the latent process $\bx_i$, the marginal likelihood is given by
\begin{eqnarray}
\label{eq:likelihood}
\prod_{i=1}^n\left[\int f(\by_i | \bx_i, \bt_i, \bSigma_i) f(\bxi_i|\bD_0) \left\{\prod_{j=1}^J f(\bzeta_{ij}|\bD_1)\right\} f(T_i, \Delta_i | h_0, \bz_i, \bx_i,\bgamma_z, \bgamma_\eta) d\bbbeta_i \right]. 
\end{eqnarray}

\section{Model Estimation}
\label{sec:estimation}
\subsection{Two-Step Method}
\label{sec:two-step}
A naive estimation method would be to use a two-step method by first predicting the scores from the longitudinal data model \eqref{eq:model_long} and then doing plug-in for Cox regression model \eqref{eq:model_cox}.
The first step is non-trivial so we shall provide some details.

First, for each longitudinal biomarker, the  mean function is estimated by penalized splines \citep{Eilers:96}. Next, we adopt the fast covariance estimation method for multivariate sparse functional data  in \cite{li2018fast} to obtain estimates of the auto- and cross-covariance functions, $\widehat{\C}_{jj^{\prime}}$, and error variances, $\widehat{\sigma}_j^2$, via bivariate penalized splines. Finally, we treat the estimates as true auto- and cross-covariances and estimate $\beta_j$ as in Web Appendix A. Once $\widehat{\beta}_j$ are obtained, $\widehat{\C}_0$ can be solved by least squares using equation \eqref{eq:moment:generic} with $j \neq j'$. Then, $\widehat{\C}_1$ can be solved similarly  using the same equation with $j = j'$. The negative eigenvalues will be discarded to ensure that the covariances are positive semi-definite.
We then use the conditional expectation approach for predicting the scores, a popular approach in traditional joint modeling \citep{wulfsohn1997joint} and sparse functional data analysis \citep{Yao:05}; see Web Appendix C for details.
Finally, the predicted scores, $\widehat{\E}(\bbbeta_i | \by_i)$, where the estimates of fixed quantities are plugged in, will be used in the Cox regression.

Despite its computational advantage, the two-step method has some well-known drawbacks: (1) it is a marginal approach which ignores the inherent correlation between the longitudinal and survival process and often leads to inferior statistical efficiency; (2) the predicted scores in the first step are usually biased and the estimation error will propagate into the subsequent Cox regression. Nevertheless, we shall compare the two-step method with the proposed estimation method below and demonstrate the superiority of the latter one in the numerical study. In addition, the estimates from the two-step method can be used as initial values for the joint estimation method.

\subsection{Monte Carlo EM Method}
\subsubsection{Reduced Rank Splines}\label{sec:reduced}
Following \cite{yao2007functional} and \cite{huang2014joint}, we use reduced rank splines for modeling the smooth mean functions and covariance functions.
Let $\bb(t) = \{B_1(\cdot), \cdots, B_c(\cdot)\}^{\top}$ be the vector of B-spline basis functions in the unit interval \citep{deBoor:78}, where $c$ is the number of equally-spaced interior knots plus the order (degree plus $1$) of the B-splines. 
We model the mean function $\mu_j(t)$ by $\bb(t)^{\top}\balpha_j$,
where $\balpha_j$ is the coefficient vector of the $j$th mean function.
Let $\bG = \int \bb(t) \bb(t)^{\top} dt \in \real^{c \times c}$, which is positive definite \citep{zhou1998local}. Then $\widetilde{\bb}(t) = \bG^{-\frac{1}{2}}\bb(t)$ are  orthonormal B-spline bases.
For the covariance functions $\mathcal{C}_0$ and $\mathcal{C}_1$, we approximate their $\ell$th eigenfunctions $\phi_\ell(t)$ and $\psi_\ell(t)$ by $\widetilde{\bb}(t)^{\top}\btheta_{0\ell}$ and $\widetilde{\bb}(t)^{\top}\btheta_{1\ell}$, respectively, where $\btheta_{0\ell}$ and $\btheta_{1\ell}$ are coefficient vectors.
Let $\bTheta_0 = [\btheta_{01},\ldots, \btheta_{0L_0}]$
and $\bTheta_1 = [\btheta_{11},\ldots, \btheta_{1L_1}]$.
Then the orthonormality of eigenfunctions gives the constraints,
$\bTheta_0^{\top} \bTheta_0 = \bI_{L_0 \times L_0}$ and $\bTheta_1^{\top} \bTheta_1 = \bI_{L_1 \times L_1}$.
These constraints are equivalent to $\btheta_{0 \ell}^{\top} \btheta_{0 \ell^{\prime}} = \btheta_{1 \ell}^{\top} \btheta_{1 \ell^{\prime}} = 1_{\{\ell = \ell^{\prime}\}}$.

\subsubsection{E-Step}
Although nonparametric  functions are components of the proposed model,
their spline representations allow a parametric estimation
based on the EM algorithm.  
The full data likelihood depends on the latent random variables $\bbbeta_i$ and can be optimized via the EM method, which treats $\bbbeta_i$ as missing values and iterates between E-steps and M-steps until convergence. Such a strategy is often deployed in parametric joint modeling \citep{wulfsohn1997joint}. 
We shall use the Monte Carlo EM algorithm, an alternative to the Gaussian-Hermite quadrature,  to approximate the numerical integrals in the E-step. 

Let $\balpha = (\balpha_1^{\top}, \cdots, \balpha_J^{\top})^{\top}$ be the vector of spline coefficients for the mean functions,
$\bbeta = {(\beta_1, \cdots, \beta_J)}^{\top}$ the vector of scaling parameters and $\bsigma^2 = (\sigma_1^2, \cdots, \sigma_J^2)^{\top}$ the vector of error variances.
Denote by $\bOmega = \{h_0, \bbeta,\bgamma_z, \bgamma_\eta, \bD_0, \bD_1, \balpha, \bTheta_0, \bTheta_1, \bsigma^2\}$ the set of parameters and  $\widehat{\bOmega} = \{\widehat{h}_0, \widehat{\bbeta}, \widehat{\bgamma}_z,  \widehat{\bgamma}_\eta, \widehat{\bD}_0, \widehat{\bD}_1, \widehat{\balpha}, \widehat{\bTheta}_0, \widehat{\bTheta}_1, \widehat{\bsigma}^2\}$ the estimate. 
Let $g(\cdot)$ be any smooth function of $\bbbeta_i$,  then the conditional expectation $\E\{g(\bbbeta_i) | T_i, \Delta_i, \bz_i, \by_i, \bt_i, \widehat{\bOmega} \} $ is given by
\begin{eqnarray*}
	\frac{\int g(\bbbeta_i) f(T_i, \Delta_i | \widehat{h}_0, \bz_i, \bbbeta_i,  \widehat{\bgamma}_z,  \widehat{\bgamma}_\eta)  f(\bbbeta_i| \by_i, \bt_i, \widehat{\balpha}, \widehat{\bbeta}, \widehat{\bTheta}_0, \widehat{\bTheta}_1, \widehat{\bD}_0, \widehat{\bD}_1, \widehat{\bsigma}^2)  d\bbbeta_i}{\int f(T_i, \Delta_i | \widehat{h}_0, \bz_i, \bbbeta_i,  \widehat{\bgamma}_z, \widehat{\bgamma}_\eta)  f(\bbbeta_i| \by_i, \bt_i, \widehat{\balpha}, \widehat{\bbeta}, \widehat{\bTheta}_0, \widehat{\bTheta}_1, \widehat{\bD}_0, \widehat{\bD}_1, \widehat{\bsigma}^2)  d\bbbeta_i},
\end{eqnarray*}
where $f(T_i, \Delta_i | \widehat{h}_0, \bz_i, \bbbeta_i, \widehat{\bgamma}_z, \widehat{\bgamma}_\eta)$ is the conditional likelihood in \eqref{eq:lcox}, and the second part of the denominator
can be obtained from the joint normality of $\bbbeta_i$ and $\by_i$,  given the data and parameter estimates; see Web Appendix C.
We now use $\E_i\{g(\bbbeta_i)\}$ to denote the conditional expectation for convenience.
In the E-step, because the integrals for the conditional expectations have no closed form solution,  we use Monte Carlo approximation 
\begin{eqnarray*}
	\E_i\{g(\bbbeta_i)\} \approx \frac{\sum_{q=1}^Q g(\bbbeta_i^{(q)}) f(T_i, \Delta_i | \widehat{h}_0, \bz_i, \bbbeta_i^{(q)}, \widehat{\bgamma}_z, \widehat{\bgamma}_\eta)}{\sum_{q=1}^Q f(T_i, \Delta_i | \widehat{h}_0, \bz_i, \bbbeta_i^{(q)}, \widehat{\bgamma}_z, \widehat{\bgamma}_\eta)},
\end{eqnarray*}
where $\bbbeta_i^{(q)}$ is the $q$th sample from the normal distribution $f(\bbbeta_i| \by_i, \bt_i, \widehat{\balpha}, \widehat{\bbeta}, \widehat{\bTheta}_0, \widehat{\bTheta}_1, \widehat{\bD}_0, \widehat{\bD}_1, \widehat{\bsigma}^2)$, and $Q$ random samples are drawn. 
To accelerate the convergence, we use the estimates from the two-step method as the initial values of the parameters. 

\subsubsection{M-Step}
Estimates of the current iteration can be obtained by optimizing separate parts of the joint likelihood \eqref{eq:likelihood} in the M-step, because each part only involves disjoint sets of parameters. 
Specifically, $\balpha$, $\bbeta$, $\bTheta_0$, $\bTheta_1$ and $\bsigma^2$ can be estimated iteratively by minimizing the expected negative log likelihood of the longitudinal process \eqref{eq:llong},
\[
\sum_{i=1}^n \sum_{j=1}^J\left[\frac{m_i}{2}\log(2\pi\sigma_j^2) + \frac{1}{2\sigma_j^2}\E_i \left\{\by_{ij} - \bB_i\balpha_j - \beta_j\left( \widetilde{\bB}_i\bTheta_0\bxi_i + \widetilde{\bB}_i\bTheta_1\bzeta_{ij}\right) \right\}^2 \right].
\]
We adopt an iterative algorithm to cyclically estimate the columns of $\bTheta_0$ and $\bTheta_1$, and deploy an ad hoc step to satisfy the orthonormality constraints on the parameter matrices. The parameters in the diagonal matrices
$\bD_0$ and $\bD_1$ are estimated by minimizing the expected negative logarithm of $\prod_{i=1}^n f(\bxi_i | \bD_0)$ and $\prod_{i=1}^n\prod_{j=1}^J f(\bzeta_{ij} | \bD_1)$, respectively.
The baseline hazard function
$h_0$ and  the parameter vectors $\bgamma_z$ and $\bgamma_\eta$ in the Cox regression can be estimated according to the expected negative log likelihood of the survival process \eqref{eq:lcox},
\[
\sum_{i=1}^n\left[-\Delta_i\left\{\log h_0(T_i) + \bz_i^{\top}\bgamma_z +  \E_i(\beeta_i^{\top}\bgamma_\eta)\right\} + \int_0^{T_i} h_0(u) \E_i\left\{\exp\left(\bz_i^{\top}\bgamma_z + \bbbeta_i^{\top}\bgamma_\eta\right)\right\} du \right].
\]
In particular, the baseline hazard $h_0$ is estimated nonparametrically by the Breslow estimator, and $\bgamma_z$ and $\bgamma_\eta$ are  updated by a one-step Newton-Raphson algorithm inside the loop.
The estimated standard errors of the Cox regression coefficients can be obtained by inverting the observed information matrix.
We defer the technical details to Web Appendix D.

\section{Model Selection}
\label{sec:mod_sel}
As described in Section 3.2, cubic B-splines are used for approximating  mean functions and eigenfunctions. We use equally-spaced knots for constructing the splines and for simplicity, we use the same number of knots for all spline functions. 
Following \cite{huang2014joint}, we use the asymptotic theory in \cite{li2010uniform} to determine the number of basis functions according to the sample size. For the simulation and data application, we use 9 spline bases ($c = 9$) which is found to work well. See Web Appendix E for implementation details.

The number of eigenfunctions is an important tuning parameter since it determines the functional characteristics of the latent stochastic process. We use information criteria for model selection, which requires an evaluation of the model complexity, the degrees of freedom of the model.
It can be shown that the negative log likelihood is given by
$\ell_n = -2  \sum_{i=1}^n \left[ \log f(\by_i | \widehat{\bOmega}) + \log \E \{f(T_i, \Delta_i | \bz_i, \bbbeta_i, \widehat{\bOmega})\} \right]$,
where $f(\by_i | \widehat{\bOmega})$ is a normal density with the covariance described in Web Appendix C, and the expectation can be approximated by $Q^{-1}\sum_{q=1}^Q f(T_i, \Delta_i | \widehat{h}_0, \bz_i, \bbbeta_i^{(q)}, \widehat{\bgamma}_z, \widehat{\bgamma}_\eta)$.

We approximate the degrees of freedom via the number of effective parameters, 
\begin{eqnarray*}
	\textnormal{df} := Jc + (L_0 + L_1)(c+1) + P + L_0 + J L_1 + 2J - 1 - \frac{L_0(L_0 + 1)}{2} - \frac{L_1(L_1 + 1)}{2},
\end{eqnarray*}
where $Jc$ is  the number of parameters for estimating the mean functions, $(L_0 + L_1)(c+1)$ is corresponding to the eigen pairs, $P + L_0 + J L_1$ is the number of coefficients in Cox regression, $2J - 1$ is corresponding to the error variance $\sigma_j^2$ and the scaling factor $\beta_j$, and the last two terms are due to  orthonormality constraints on $\bTheta_0$ and $\bTheta_1$, the matrix of spline coefficients for eigenfunctions (see Section \ref{sec:reduced}). 
Therefore, we may calculate
$\textnormal{AIC} =  \ell_n + 2 \cdot \textnormal{df}$, and $\textnormal{BIC} =  \ell_n + \log n \cdot \textnormal{df}$.
We shall use a two-dimensional grid for selecting the two tuning parameters $L_0$ and $L_1$.

\section{Data Analysis}
\label{sec:application}
We apply the proposed functional joint model (denoted as FJM) to the ADNI data for 
jointly characterizing the varying patterns of the multivariate longitudinal outcomes and their association with time to diagnosis of AD.
The data are from the first two phases of ADNI, which contain $803$ participants with amnestic mild cognitive impairment (MCI, a transition risk state between normal state and AD state) at baseline who had at least one follow-up visit. 
Participants of the first phase were scheduled to be assessed at baseline, $6$, $12$, $18$, $24$, and $36$ months with additional annual follow-ups included in the second phase. Note that the exact follow-up times can actually vary.
Thus, for the combined data, the average number of visits is $4.72$.
For the analysis,  the following variables are used as baseline covariates: baseline age (mean: $74.4$, standard deviation: $7.3$, range $55.1-89.3$), gender ($36.1\%$ female), years of education (mean: $15.6$, standard deviation: $3.0$, range $4-20$), and the number of apolipoprotein E $\epsilon$4 alleles (APOE4, $56\% >= 1$), given their potential effects on AD progression \citep{fleisher2007clinical}.

We consider various models listed in Table \ref{table:models}, including the proposed functional joint model and its  variants.
The hazard model is specified as
\begin{eqnarray*}
	h_i(t) = h_0(t) \exp\left\{\textnormal{Age}_i \gamma_a + \textnormal{Gender}_i \gamma_g + \textnormal{Education}_i \gamma_e + \textnormal{APOE4}_i \gamma_{\epsilon} + \F(\bx_i, t)\right\}
\end{eqnarray*}
and the form of $\F(\bx_i, t)$ is given in Table \ref{table:models}.
Reduced models A and B share the same MFMM submodel \eqref{eq:model_long} as FJM.
Reduced model A is a special case of the proposed model with only shared components $\bxi_i$ contributing to survival risk.
Reduced model B is another special case with only outcome-specific components $\bzeta_{ij}$ contributing to survival risk.
Compared to FJM,
reduced model C  only has shared components $\bxi_i$ in both submodels.
To fit FJM, we use the settings of splines described in Section \ref{sec:mod_sel}.
The number of Monte Carlo samples and stopping criteria are set as described in Web Appendix E. 
We select the numbers of eigenfunctions according to BIC as we shall show in Section \ref{sec:sim} that it performs well for model selection.
The three reduced models can be estimated similarly as the full FJM using the proposed MCEM approach and model selection can be similarly carried out using BIC.
In addition, we consider the parametric
multivariate joint linear model (denoted as MJM) proposed by \cite{henderson2000joint} and implemented in \texttt{R} package \texttt{joineRML} \citep{hickey2018joinerml}.
With slight abuse of notation, we denote by $\beta_{0j}$ and $\beta_{1j}$ the fixed effects, and $b_{0ij}$ and $b_{1ij}$ the random intercept and slope in a linear mixed effects model for the $j$th outcome. For MJM, the corresponding Cox coefficient for the $j$th outcome is $\gamma_j$.

Table \ref{table:models} presents the overall performance of the various models. 
First,  FJM has the highest likelihood and smallest AIC and BIC, which compares favorably against the other models. 
Second, Reduced model A is the closest to FJM in terms of the three criteria, followed by Reduced model B and Reduced model C. 
We shall see later that the shared components play a major role in determining AD risk, and hence Reduced model A outperforms Reduced model B. Reduced model C not only overlooks the outcome-specific components in Cox regression, but also ignores that heterogeneity in modeling longitudinal outcomes, which explains why its performance is inferior to Reduced models A and B. 
Furthermore, MJM is outperformed by all other models,
indicating that it gains to model the longitudinal outcomes nonlinearly. 
Finally, we include the concordance index \citep{harrell2015regression} for evaluating the predictive ability of survival models as an additional criterion.
Again, FJM has the highest concordance index, while other models are slightly inferior to it. It is not surprising to see that Reduced model B is ranked last among the competitors since the shared components are primary contributors to hazard risk.

As suggested by one reviewer, a partial functional linear model (PFL) might be adopted proposed if the survival part is the primary interest \citep{kong2016partially}.
PFL treats the functional outcomes as cross-sectional covariates and uses a linear combination of eigen scores of each functional variable as the predictor in Cox regression. Despite the potential multicollinearity of the scores, this model is similar to using $W_{ij}(t)$ in MFMM but does not separate the shared component from the outcome-specific components.
One  primary objective of joint modeling is to understand the associations between features of the longitudinal outcomes and time to disease progression \citep{tsiatis2004joint}, and it is known that multicollinearity may be an issue for this objective.
Nonetheless, we have compared MFMM, MFPCA, and PFL for survival prediction and found that
MFMM performs best in terms of the concordance index and PFL is outperformed by MFPCA. 
The above models are also compared for fitting the longitudinal outcomes further showing advantages of MFMM;
see Web Appendix F for details.

\begin{table}[hbp]
	\caption{\label{table:models} Model comparison. The ``best model" row gives selected number(s) of eigenfunctions by BIC.}
	\centering
	\scalebox{0.70}{
		\begin{tabular}{cccccc}
			\hline
			& FJM & Reduced A & Reduced B & Reduced C & MJM \\
			\hline
			$X_{ij}(t)$ & MFMM & MFMM & MFMM & $\mu_j(t) + \beta_j U_{i}(t)$ & $\beta_{0j} + \beta_{1j} t + b_{0ij} + b_{1ij} t$\\
			$\F(\bx_i, t)$ & $\sum_{\ell=1}^{L_0}\xi_{i\ell}\gamma_{0 \ell} + \sum_{j=1}^{J}\sum_{\ell=1}^{L_1}\zeta_{i j \ell}\gamma_{1 j \ell}$ & $\sum_{\ell=1}^{L_0}\xi_{i\ell}\gamma_{0 \ell}$ & $\sum_{j=1}^{J}\sum_{\ell=1}^{L_1}\zeta_{i j \ell}\gamma_{1 j \ell}$ & $\sum_{\ell=1}^{L_0}\xi_{i\ell}\gamma_{0 \ell}$ & $\sum_{j=1}^J (b_{0ij} + b_{1ij} t)\gamma_j$ \\
			Best model & $(L_0, L_1) = $ (2, 2) & $(L_0, L_1) = $ (2, 2) & $(L_0, L_1) = $ (2, 2) & $L_0 = 2$ & NaN \\
			log likelihood & -37773.30 & -37849.41 & -37954.25 & -39718.01 & -47249.77 \\
			AIC & 75754.60 & 75906.83 & 76116.50 & 79590.03 & 94657.54 \\
			BIC & 76242.19 & 76394.42 & 76604.09 & 79951.03 & 95272.71 \\
			Concordance & 0.86 & 0.85 & 0.77 & 0.84 & 0.85 \\
			\hline
		\end{tabular}
	}
\end{table}

Table \ref{table:cox coeff} summarizes the estimated Cox coefficients from the functional joint model.
We have the following remarks.
(1) The results show that APOE4 is significantly associated with AD risk at level $0.05$, which is consistent with existing  AD studies. 
In particular, the presence of APOE4 allele increases the hazard of AD diagnosis by $39.10\%$ while adjusting for other covariates. 
(2) The parameters $\gamma_{01}$, $\gamma_{02}$ capture the effects of the latent disease process $U_i(t)$ manifested by the five biomarkers. The significance of these effects indicates the  contribution of the latent profile shared among the longitudinal outcomes to the hazard of AD conversion, after adjusting for baseline clinical covariates. The result agrees with the excellent predictive performance of the five biomarker reported in \cite{li2017prediction}.
(3) The proposed functional joint model sheds new insight on AD study by successfully identifying important  associations between individual longitudinal outcomes and the survival. 
In Table \ref{table:cox coeff}, the individual effects of RAVLT-immediate $\gamma_{21}$, $\gamma_{22}$ are significant, while others are not.
These results suggest that the progression patterns of the longitudinal outcomes (ADAS-Cog 13, RAVLT-learn, MMSE, and FAQ) contribute to AD diagnosis mainly through the shared latent profile, not through their outcome-specific progression. By contrast, RAVLT-immediate contributes through the outcome-specific progression in addition to the shared latent progression.
Our findings are again supported by an independent study \citep{li2019penalized}, which applied a penalized method and consistently selected RAVLT-immediate as the only significant risk factor of AD conversion.
\begin{table}[htp]
	\caption{\label{table:cox coeff} Estimates (standard errors) of Cox regression coefficients from functional joint model. An asterisks indicates significance at level 0.05.}
	\centering
	\scalebox{0.9}{
		\begin{tabular}{cccc}
			\hline
			FJM & Coefficient & Estimate (standard error) & P-value \\
			\hline
			Age & $\gamma_a$ & -0.01 (0.01) & 0.06 \\
			Gender (Female) & $\gamma_g$ & 0.27 (0.23) & 0.25  \\
			Education & $\gamma_e$ & 0.03 (0.03) & 0.29 \\
			APOE4 & $\gamma_{\epsilon}$ & 0.33 (0.15)$^*$ & 0.03 \\
			\multirow{2}{*}{Shared latent progression} &  $\gamma_{01}$ & 0.33 (0.02)$^*$ & $3e - 51$ \\
			\multirow{2}{*}{} & $\gamma_{02}$ & 0.29 (0.09)$^*$ & $0.01$ \\
			\multirow{2}{*}{ADAS-Cog 13 progression} & $\gamma_{11}$ & 0.00 (0.08) & 0.97 \\
			\multirow{2}{*}{} & $\gamma_{12}$ & -0.24 (0.22) & 0.28 \\
			\multirow{2}{*}{RAVLT-immediate progression} & $\gamma_{21}$ & 0.20 (0.08)$^*$ & 0.01 \\
			\multirow{2}{*}{} & $\gamma_{22}$ & 0.73 (0.19)$^*$ & $9e - 5$ \\
			\multirow{2}{*}{RAVLT-learn progression} & $\gamma_{31}$ & 0.03 (0.08) & 0.74 \\
			\multirow{2}{*}{} & $\gamma_{32}$ & -0.06 (0.29) & 0.83 \\
			\multirow{2}{*}{MMSE progression} & $\gamma_{41}$ & -0.04 (0.06) & 0.51 \\
			\multirow{2}{*}{} & $\gamma_{42}$ & -0.26 (0.28) & 0.35 \\
			\multirow{2}{*}{FAQ progression} & $\gamma_{51}$ & 0.09 (0.07) & 0.16 \\
			\multirow{2}{*}{} & $\gamma_{52}$ & -0.21 (0.20) & 0.29 \\
			\hline
		\end{tabular}
	}
\end{table}

\begin{figure}[htp]
	\centering
	\scalebox{0.4}{
		\includegraphics[]{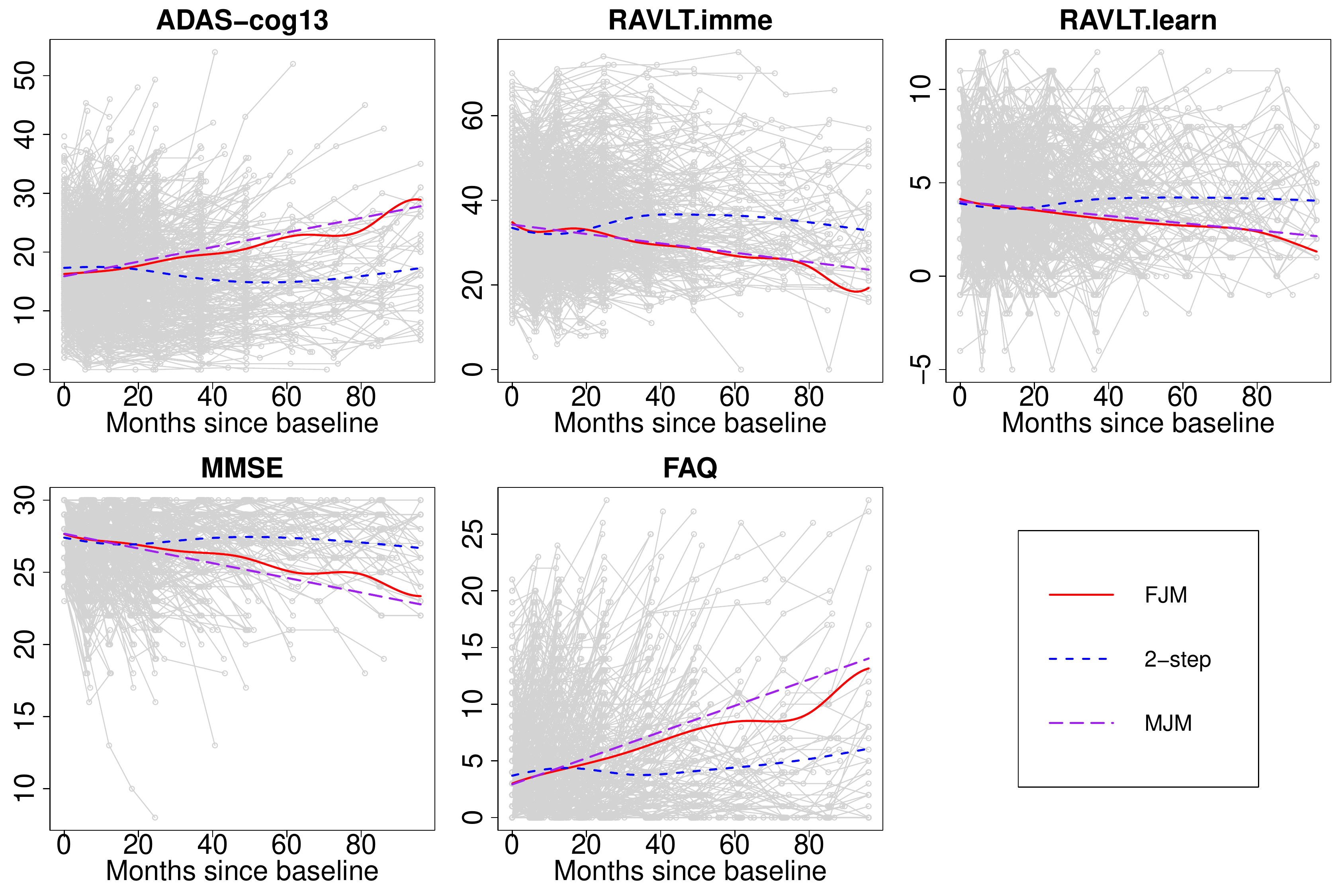}
	}
	\caption{\label{figure:mean} Estimated mean functions. Gray lines: longitudinal outcomes.}
\end{figure}

Figure \ref{figure:mean} presents the estimated mean functions of longitudinal outcomes by three methods.
Both FJM and MJM are based on joint estimation, and show a similar trend: the mean curves are progressing toward mental deterioration over the months, suggesting an increased risk of developing AD.
These findings confirm the intuition since the participants in the study suffer from MCI, which causes cognitive decline toward dementia.
Moreover, FJM can further characterize the nonlinear pattern of the biomarkers. 
Although MJM only provides linear estimates, it correctly identifies the deteriorating trend. 
The two-step method (denoted as 2-step) fails in capturing such a degenerate trend since the curves are relatively stable during the study period; this might be because the estimates are biased due to its marginal nature. 

Additional results for the ADNI data are presented in Web Appendix F.

\section{Simulations}
\label{sec:sim}

\subsection{Simulation Settings}
In this section, we compare the performance of the joint estimation, and the two-step method, of the proposed model.
We consider two cases of data generation and replicate each for $100$ times. Here, we focus on case 1 which is a realistic setting with data generated according to the fitted model of the ADNI study in Web Appendix G. See Web Appendix H for case 2 which is an alternative setting, but its results show a similar pattern as case 1.

The longitudinal data are generated according to MFMM \eqref{eq:model_long} with two outcomes $J = 2$.
The outcome-specific mean functions, scaling parameters, and error variances are derived from the estimates of ADAS-Cog 13 and RAVLT-immediate.
We set two principal components for both two covariances, and the eigenfunctions are specified as the estimates of $\C_0(s, t)$ and $\C_1(s, t)$.
The eigen scores $\xi_{i \ell}$'s are generated from a normal distribution $\N(0, d_{0 \ell})$ with $d_{0 1} = 95.41$ and $d_{0 2} = 5.04$.
The outcome-specific eigen scores $\zeta_{i j \ell}$'s are generated similarly with $d_{1 1} = 21.90$ and $d_{1 2} = 2.05$. 
We set the scaling parameters $\beta_1 = 1$ and $\beta_2 = -1.44$. 
The white noise $\epsilon_{ijk}$s are sampled from a normal distribution $\N(0, \sigma_j^2)$, where $\sigma_1^2 = 9.49$ and $\sigma_2^2 = 21.98$. The observed time points $t_{ijk} = t_{ik}$ are $11$ fixed time points of the ADNI study mapped to the interval $[0,1]$.

The time-to-event data are generated according to Cox regression \eqref{eq:model_cox} with the coefficients set as the estimates of the common components and outcome-specific components of ADAS-Cog 13 and RAVLT-immediate. We use the baseline hazard function $h_0(t) = 1$, and specify the linear hazard rate function as $ \sum_{\ell=1}^{2}\xi_{i\ell}\gamma_{0 \ell} + \sum_{j=1}^{2}\sum_{\ell=1}^{2}\zeta_{i j \ell}\gamma_{1 j \ell}$, 
where the Cox coefficients are $\bgamma_0 = (0.33, 0.31)^{\top}$, $\bgamma_{11} = (0.01, -0.27)^{\top}$, $\bgamma_{12} = (0.25, 0.80)^{\top}$. 
Then failure times are drawn independently from a standard exponential distribution.
Censoring times $C_i$s are generated independently from a uniform distribution on $[0, c_0]$, where $c_0$ is a constant and the final truncation time $\tau = 1$ is used so that the censoring rate is around $65\%$. 
For each subject, only measurements at $t_{ik} \leq T_i$ are retained. We generate data with $803$ subjects, and the average number of observations per subject is around $5.5$. All of these settings are close to the real case of the ADNI study.

\subsection{Simulation Results}
\begin{figure}[htp]
	\centering
	\scalebox{0.5}{
		\includegraphics[]{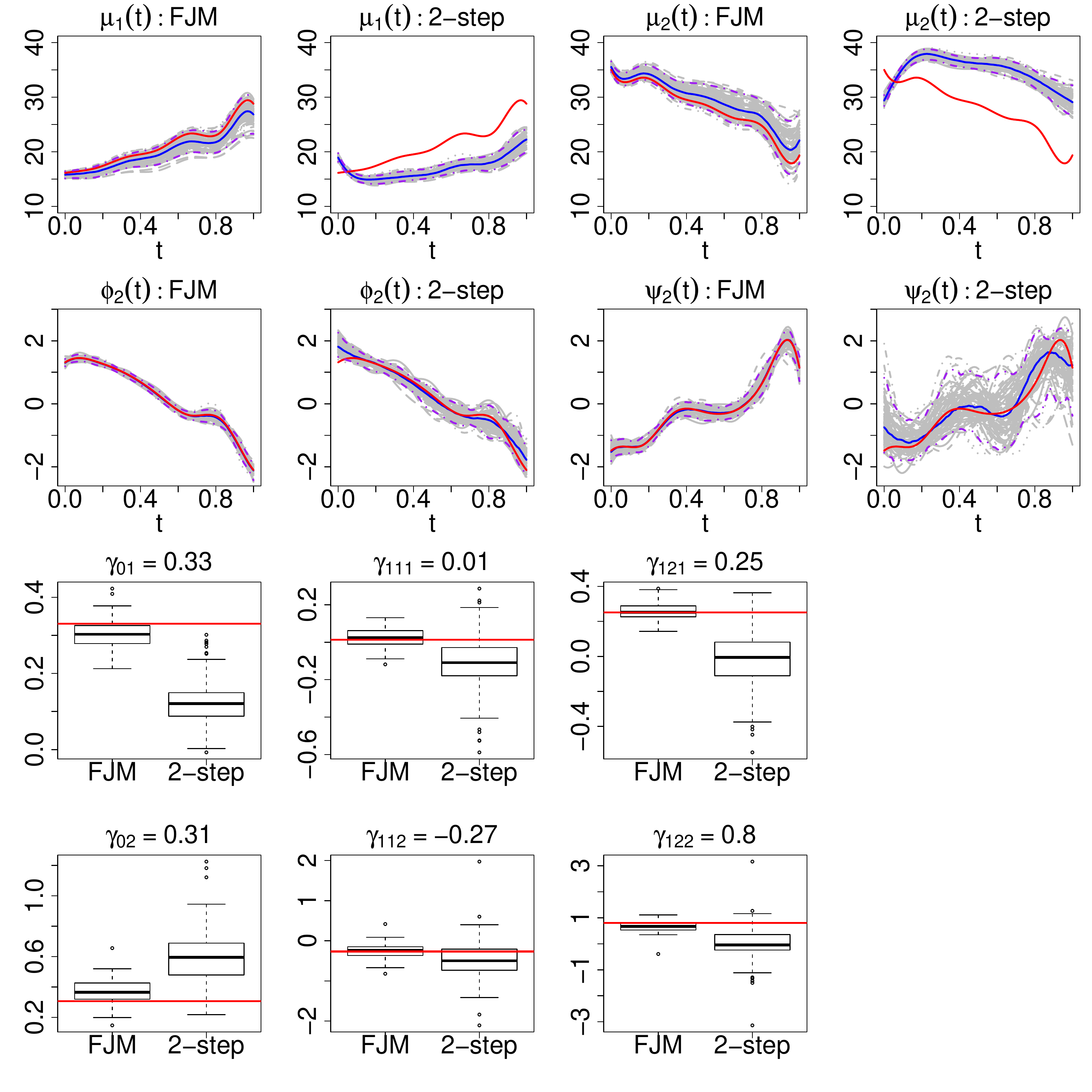}
	}
	\caption{\label{fig:sim} Estimated functions/parameters of $100$ replications.
		The first row: estimated mean functions;
		the second row: estimated eigenfunctions;
		the last two rows: estimated Cox coefficients. 
		Red lines: true functions/parameters; gray lines: estimated functions; blue lines: medians of estimates; dashed purple lines: $95\%$ point-wise confidence bands.}
\end{figure}

We use the settings of model fitting described in Web Appendix E.
First, we fix the number of principal components as the truth $L_0 = L_1 = 2$ and estimate model components.
In most of the replications, FJM converges within $200$ iterations. 
The first two rows of Figure \ref{fig:sim} present the estimated mean functions and eigenfunctions for $\phi_2(t)$ and $\psi_2(t)$. While the medians of FJM are close to the truth, the two-step method has significant bias over the time. Furthermore, we obtain the point-wise confidence bands based on the quantiles of all the replications. The $95\%$ confidence bands of FJM are able to cover the truth, but this is not the case for the two-step method since the true means lie outside its $95\%$ confidence bands.
The last two rows of Figure \ref{fig:sim} summarize the estimates of the Cox coefficients. FJM is reasonably close to the truth, but the two-step method shows significant bias.

Finally, we use AIC and BIC to select the number of eigenfunctions in the covariances and evaluate the performance of the proposed approaches. 
One may define the candidate ranks of the covariance by the proportion of variance explained (PVE) in the marginal MFMM stage since the two-step method is used for providing initial values.
For the two-step method, the number of principal components can be selected by using either AIC or BIC solely based on Cox regression as in \cite{kong2018flcrm}. 
For the two-step method, the rates of correctly selecting two principal components for the covariances are $0.40$ and $0.38$ using BIC, respectively.   
By contrast, the proposed approach for FJM achieves excellence in practice, the correct selection rates are $1.00$ for all using BIC. 
For both the two methods, the rates of AIC are slightly lower than those of BIC, so we use BIC for rank selection in the data application.

In summary, FJM shows very competitive performance and is superior to the two-step method in terms of estimation and rank selection in all scenarios. Additional simulation results are included in Web Appendix H.

\section{Discussion}
\label{sec:discussion}
Our work can be extended in several directions.
First, the MFMM framework is flexible to further account for heterogeneity across multiple longitudinal outcomes. For example, one might use two different scaling parameters multiplying $U_i(t)$ and $W_{ij}(t)$ in model \eqref{eq:model_long}.
Moreover, one might model $W_{ij}(t)$ with heterogeneous covariances to incorporate any prior information.
We have adopted model \eqref{eq:model_long} with homogenous covariances in this paper as it is found to better fit the ADNI data.
Second, it is worth developing a joint integrative modeling framework to incorporate imaging, genetic, and longitudinal biomarkers into the Cox regression \eqref{eq:model_cox} and comparing it with a predictive model proposed by \cite{kong2015predicting}, which treats time to AD as the survival outcomes and uses multimodal data to predict AD progression. 
Finally, the theoretic properties of the proposed joint model are unclear, which warrant avenues for future research.

\section*{Appendices}
Web Appendix A includes the proof for identifiability of MFMM Model and details of the correlation between outcomes.
Web Appendix B presents the connection between MFMM and MFPCA.
Web Appendix C describes score prediction for the two-step method.
Web Appendix D describes technical details of the M-step.
Web Appendix E describes implementation details.
Web Appendix F presents additional results in Section 5.
Web Appendix G contains a sensitivity analysis for the ADNI data.
Web Appendix H includes additional simulation results in Section 6.

\subsection*{Web Appendix A: Mathematical Details of MFMM Model}
\subsubsection*{Identifiability of MFMM Model}
\label{sec:identi}
We show that $\beta_j$, $\C_0$ and $\C_1$ can be uniquely identified. 
Note that the covariance functions $\C_{jj'}(s,t)$ are well defined.
First consider $J \geq 3$. To determine $\beta_j$ with $j\neq 1$ consider $j'\neq j$ and $j'\neq 1$. 
Then $\C_{j'1}(s,t) = \beta_{j'} \C_0(s,t)$ and $\C_{jj'}(s,t) = \beta_j\beta_j'\C_0(s,t)$. 
It follows that $\C_{jj'}(s,t) = \beta_j \C_{j'1}(s,t)$ and $\beta_j$ can be solved by least squares and the solution is unique. Then, by $\C_{jj'}(s,t) = \beta_j\beta_j'\C_0(s,t)$, we conduct least squares again to uniquely determine $\C_0$. Similarly, by equation (2), we obtain $\C_1$. 
Now consider $J=2$. We have three equations:
$\C_{11}(s,t) = \C_0(s,t) + \C_1(s,t)$,
$\C_{22}(s,t) = \beta_2^2 \C_0(s,t) + \beta_2^2 \C_1(s,t)$,
$\C_{12}(s,t) = \beta_2 \C_0(s,t)$.
Then, we have $\C_{22}(s,t) / \C_{11}(s,t) = \beta_2^2$ and we can solve $\beta_2$ as before and determine its sign according to the equation $\C_{12}(s,t) = \beta_2 \C_0(s,t)$ since $\C_0(s,t)$ is a covariance function and its diagonal elements should be positive. 
Then $\C_0$ can be solved by the third equation and finally $\C_1$ can be solved by the first two equations.

\subsubsection*{Correlation between Outcomes}
For $j \neq j'$, we have $\C_{jj'}(s,t) = \Cov\{X_{ij}(s), X_{ij'}(t)\} = \beta_j \beta_j' \C_0(s,t)$ according to Equation (2). 
Recall that $\C_0(s,t) = \Cov\{U_i(s), U_i(t)\}$, so its diagonal elements are variances (positive). Therefore, if $\beta_j\beta_j' < 0$, the two outcomes are negatively correlated; otherwise, they are positively correlated. 

\subsection*{Web Appendix B: Connection between MFMM and MFPCA}
Denote by $\bx_i(t) = (X_{i1}(t), \cdots, X_{iJ}(t))^{\top}$ the $J$ longitudinal outcomes evaluated at time $t$ for the $i$th subject, and let other notations follow those in Section 2.3, then we have 
\begin{eqnarray*}
	& & \Cov(\bx_i(t)) = \begin{pmatrix} \beta_1^2 \bPhi(t)^{\top} \bD_0 \bPhi(t) + \beta_1^2 \bPsi(t)^{\top} \bD_1 \bPsi(t) &  \ldots & \beta_1\beta_J \bPhi(t)^{\top} \bD_0 \bPhi(t) \\
		\vdots & \ddots & \vdots \\
		\beta_J \beta_1 \bPhi(t)^{\top}\bD_0 \bPhi(t) & \ldots & \beta_J^2 \bPhi(t)^{\top} \bD_0 \bPhi(t) + \beta_J^2 \bPsi(t)^{\top} \bD_1 \bPsi(t)
	\end{pmatrix} \\
	& & = \begin{pmatrix} \beta_1^2 \bPhi(t)^{\top}  \bD_0 \bPhi(t)&  \ldots & \beta_1\beta_J \bPhi(t)^{\top}\bD_0 \bPhi(t) \\
		\vdots & \ddots & \vdots \\
		\beta_J \beta_1 \bPhi(t)^{\top}\bD_0 \bPhi(t) & \ldots & \beta_J^2 \bPhi(t)^{\top}  \bD_0 \bPhi(t)
	\end{pmatrix} + 
	\begin{pmatrix}  \beta_1^2 \bPsi(t)^{\top} \bD_1 \bPsi(t) &  \ldots & 0 \\
		\vdots & \ddots & \vdots \\
		0 & \ldots &  \beta_J^2 \bPsi(t)^{\top} \bD_1 \bPsi(t)
	\end{pmatrix}\\
	& & = \underbrace{\begin{pmatrix} \beta_1 \bPhi(t)^{\top} \\ \vdots \\ \beta_J\bPhi(t)^{\top} \end{pmatrix} \bD_0 \begin{pmatrix} \beta_1 \bPhi(t) & \ldots & \beta_J\bPhi(t) \end{pmatrix}}_\text{I} + \\
	& & \underbrace{\begin{pmatrix} \beta_1 \bPsi(t)^{\top} &  \ldots & \mathbf{0} \\ \vdots & \ddots & \vdots \\ \mathbf{0} & \ldots &  \beta_J^2 \bPsi(t)^{\top}\end{pmatrix} \begin{pmatrix} \bD_1 &  \ldots & \mathbf{0} \\ \vdots & \ddots & \vdots \\ \mathbf{0} & \ldots &  \bD_1 \end{pmatrix}  \begin{pmatrix} \beta_1 \bPsi(t) &  \ldots & \mathbf{0} \\ \vdots & \ddots & \vdots \\ \mathbf{0} & \ldots &  \beta_J^2 \bPsi(t)\end{pmatrix}}_\text{II}.
\end{eqnarray*}
Note that the formulations of part I and part II are similar to MFPCA and FPCA with some model simplifications, respectively. 
Therefore, the proposed MFMM can be considered as a combination of MFPCA and FPCA.

\subsection*{Web Appendix C: Score Prediction for Two-Step Method}
\label{sec:supp_scores}
Following the notation in Section 2.3,
let $\by_{ij} = \bmu_{ij} + \beta_j (\bPhi_i \bxi_i + \bPsi_i \bzeta_{ij}) + \bepsilon_{ij}$, where $\bepsilon_{ij} = (\epsilon_{ij1}, \cdots \epsilon_{ij m_i})^{\top}$ is the vector of measurement errors. 
Then, $$\by_i = \left\{\beta_1(\bPhi_i \bxi_i + \bPsi_i \bzeta_{i1})^{\top} + \bepsilon_{i1}^{\top}, \cdots, \beta_J(\bPhi_i \bxi_i + \bPsi_i \bzeta_{ij})^{\top} + \bepsilon_{iJ}^{\top}\right\}^{\top}.$$ 
We use the conditional expectation approach for predicting the scores.
We have
\[
\left(\begin{array}{c}\by_i \\\bbbeta_i\end{array}\right) \sim \mathcal{N}\left\{
\left(\begin{array}{c}\bmu_i \\ \mathbf{0}\end{array}\right),
\left(\begin{array}{cc}\Cov(\by_i)& \Cov(\by_i, \bbbeta_i)  \\\Cov(\bbbeta_i, \by_i)
& \bD  \end{array}\right) 
\right\},
\]
where $\bD = \text{blockdiag}(\bD_0,\underbrace{\bD_1,\ldots, \bD_1}_{J \text{ copies}})$,
$$
\Cov(\bbbeta_i, \by_i) = \begin{pmatrix} \beta_1 \bD_0 \bPhi_i^{\top} &  \ldots & \beta_J \bD_0 \bPhi_i^{\top} \\
\beta_1 \bD_1 \bPsi_i^{\top} & \ldots & \mathbf{0} \\
\vdots & \mathbf{0} & \vdots \\
\mathbf{0} & \ldots & \beta_J \bD_1 \bPsi_i^{\top}
\end{pmatrix},	
$$
and
$$
\Cov(\by_i) = \begin{pmatrix} \beta_1^2 \bPhi_i \bD_0 \bPhi_i^{\top} + \beta_1^2 \bPsi_i \bD_1 \bPsi_i^{\top} + \sigma_1^2\bI_{m_i}&  \ldots & \beta_1\beta_J \bPhi_i \bD_0 \bPhi_i^{\top} \\
\vdots & \ddots & \vdots \\
\beta_J \beta_1 \bPhi_i\bD_0 \bPhi_i^{\top} & \ldots & \beta_J^2 \bPhi_i \bD_0 \bPhi_i^{\top} + \beta_J^2 \bPsi_i \bD_1 \bPsi_i^{\top} + \sigma_J^2\bI_{m_i}
\end{pmatrix}.
$$
It follows that
\[
\E(\bbbeta_{i} | \by_i) = \Cov(\bbbeta_i, \by_i)\Cov(\by_i)^{-1}(\by_i - \bmu_i), \quad \Cov(\bbbeta_i|\by_i) = \bD - \Cov(\bbbeta_i, \by_i)\Cov(\by_i)^{-1}\Cov(\by_i, \bbbeta_i).
\]

\subsection*{Web Appendix D: Technical Details on M-Step}
\label{sec:m-step}
First, 
we have 
\begin{eqnarray*}
	\widehat{d}_{0 \ell} = \frac{1}{n} \sum_{i=1}^n \E_i(\xi_{i \ell}^2), \quad
	\widehat{d}_{1 \ell} = \frac{1}{nJ} \sum_{i=1}^n \sum_{j=1}^{J} \E_i(\zeta_{ij \ell}^2).
\end{eqnarray*}

Second, let $\|\cdot\|$ denote Euclidean norm, then $\widehat{\sigma}_j^2$ can be updated by
\begin{eqnarray*}
	\widehat{\sigma}_j^2 &=& \frac{1}{\sum_i {m_i}}\sum_{i=1}^n \E_i\left\{\| \by_{ij} - \widehat{\bmu}_{ij} - \widehat{\beta}_j\widehat{\bPhi}_i\bxi_i - \widehat{\beta}_j\widehat{\bPsi}_i\bzeta_{ij}  \|^2\right\} \\
	&=& \frac{1}{\sum_i {m_i}}\sum_{i=1}^n \|\by_{ij} - \widehat{\bmu}_{ij}\|^2 + \frac{\widehat{\beta}_j^2}{\sum_i {m_i}}\sum_{i=1}^n \tr\left\{\widehat{\bPhi}_i^{\top}\widehat{\bPhi}_i \E_i(\bxi_i \bxi_i^{\top})\right\} + \frac{\widehat{\beta}_j^2}{\sum_i {m_i}}\sum_{i=1}^n \tr\left\{\widehat{\bPsi}_i^{\top}\widehat{\bPsi}_i \E_i(\bzeta_{ij}  \bzeta_{ij}^{\top})\right\} \\
	&-& \frac{2 \widehat{\beta}_j}{\sum_i {m_i}}\left[\sum_{i=1}^n \left\{\by_{ij} - \widehat{\bmu}_{ij}\right\}^{\top} \widehat{\bPhi}_i \E_i\bxi_i 
	+\sum_{i=1}^n \left\{\by_{ij} - \widehat{\bmu}_{ij}\right\}^{\top} \widehat{\bPsi}_i \E_i\bzeta_{ij}\right]
	+ \frac{2 \widehat{\beta}_j^2}{\sum_i {m_i}}\sum_{i=1}^n \tr\left\{\widehat{\bPhi}_i^{\top} \widehat{\bPsi}_i \E_i (\bzeta_{ij}\bxi_i^{\top})\right\}.
\end{eqnarray*}

Third, we obtain $\widehat{\balpha}_j$ by minimizing
\begin{eqnarray*}
	\sum_{i=1}^{n} \E_i \| \by_{ij} - \widehat{\beta}_j\widehat{\bPhi}_i\bxi_i - \widehat{\beta}_j\widehat{\bPsi}_i\bzeta_{ij} - \bB_i \balpha_j\|^2,
\end{eqnarray*}
and it follows that
\begin{eqnarray*}
	\widehat{\balpha}_j = \left\{\sum_{i=1}^{n}\bB_i^{\top}\bB_i\right\}^{-1}\left\{\sum_{i=1}^{n} \bB_i^{\top} \left( \by_{ij} - \widehat{\beta}_j\widehat{\bPhi}_i\E_i \bxi_i - \widehat{\beta}_j\widehat{\bPsi}_i\E_i \bzeta_{ij}\right) \right\}.
\end{eqnarray*}

Fourth, we estimate $\bTheta_0$ through an iterative algorithm.
Given $\widehat{\btheta}_{0 \ell}$, $\ell \neq k$, $\btheta_{0 k}$ is given by minimizing
\begin{eqnarray*}
	\sum_{i=1}^{n} \sum_{j=1}^J \E_i \| \by_{ij} - \widehat{\bmu}_{ij} - \widehat{\beta}_j\widehat{\bPsi}_i\bzeta_{ij} - \widehat{\beta}_j \sum_{\ell \neq k} \widetilde{\bB}_i \widehat{\btheta}_{0 \ell}\xi_{i \ell} - \widehat{\beta}_j \widetilde{\bB}_i \btheta_{0 k}\xi_{i k}\|^2,
\end{eqnarray*}
it leads to
\begin{eqnarray*}
	& &\widehat{\btheta}_{0 k} = \left\{\sum_{i=1}^{n} \sum_{j=1}^{J} \widehat{\beta}_j^2\E_i(\xi_{ik}^2) \widetilde{\bB}_i^{\top}\widetilde{\bB}_i \right\}^{-1}\\
	& &\left[\sum_{i=1}^{n} \widetilde{\bB}_i^{\top}\sum_{j=1}^{J} \widehat{\beta}_j \left\{\E_i\xi_{ik} \left(\by_{ij} - \widehat{\bmu}_{ij}\right) - \widehat{\beta}_j\widehat{\bPsi}_i \E_i(\xi_{ik}\bzeta_{ij}) -  \widehat{\beta}_j \sum_{\ell \neq k} \widetilde{\bB}_i \widehat{\btheta}_{0 \ell} \E_i(\xi_{i k} \xi_{i \ell}) \right\} \right].
\end{eqnarray*}
Similarly, we estimate $\btheta_{1 k}$ by minimizing
\begin{eqnarray*}
	\sum_{i=1}^{n} \sum_{j=1}^J \E_i \| \by_{ij} - \widehat{\bmu}_{ij} - \widehat{\beta}_j\widehat{\bPhi}_i\bxi_i - \widehat{\beta}_j \sum_{\ell \neq k} \widetilde{\bB}_i \widehat{\btheta}_{1 \ell}\zeta_{ij \ell} - \widehat{\beta}_j \widetilde{\bB}_i \btheta_{1 k}\zeta_{ij k}\|^2,
\end{eqnarray*}
and 
\begin{eqnarray*}
	& &\widehat{\btheta}_{1 k} = \left\{\sum_{i=1}^{n} \sum_{j=1}^{J} \widehat{\beta}_j^2\E_i(\zeta_{ijk}^2) \widetilde{\bB}_i^{\top}\widetilde{\bB}_i \right\}^{-1}\\
	& &\left[\sum_{i=1}^{n} \widetilde{\bB}_i^{\top}\sum_{j=1}^{J} \widehat{\beta}_j \left\{\E_i\zeta_{ijk} \left(\by_{ij} - \widehat{\bmu}_{ij}\right) - \widehat{\beta}_j\widehat{\bPhi}_i \E_i(\zeta_{ijk}\bxi_i) -  \widehat{\beta}_j \sum_{\ell \neq k} \widetilde{\bB}_i \widehat{\btheta}_{1 \ell} \E_i(\zeta_{ijk} \zeta_{ij \ell}) \right\} \right].
\end{eqnarray*}
We repeat the procedure for each column of $\bTheta_0$ and $\bTheta_1$ until convergence. The final estimates of $\bD_0$, $\bD_1$, $\bTheta_0$ and $\bTheta_1$ are given by eigen decomposition of $\widehat{\bTheta}_0 \widehat{\bD}_0 \widehat{\bTheta}_0^{\top}$ and $\widehat{\bTheta}_1 \widehat{\bD}_1 \widehat{\bTheta}_1^{\top}$, respectively.

Next, we estimate $\beta_j$  by minimizing
\begin{eqnarray*}
	\sum_{i=1}^{n} \E_i \| \by_{ij} - \widehat{\bmu}_{ij} - \beta_j (\widehat{\bPhi}_i\bxi_i + \widehat{\bPsi}_i\bzeta_{ij})\|^2,
\end{eqnarray*}
and 
\begin{eqnarray*}
	\widehat{\beta}_j = \frac{\sum_{i=1}^{n} (\widehat{\bPhi}_i\E_i\bxi_i + \widehat{\bPsi}_i\E_i\bzeta_{ij})^{\top}(\by_{ij} - \widehat{\bmu}_{ij})  } {\sum_{i=1}^n \tr\left\{\widehat{\bPhi}_i^{\top}\widehat{\bPhi}_i \E_i(\bxi_i \bxi_i^{\top})\right\} + \sum_{i=1}^n \tr\left\{\widehat{\bPsi}_i^{\top}\widehat{\bPsi}_i \E_i(\bzeta_{ij}  \bzeta_{ij}^{\top})\right\} + 2 \sum_{i=1}^n \tr\left\{\widehat{\bPhi}_i^{\top} \widehat{\bPsi}_i \E_i (\bzeta_{ij}\bxi_i^{\top})\right\} }.
\end{eqnarray*}

The baseline hazard $h_0(t)$ can be estimated by
\begin{eqnarray*}
	\widehat{h}_0(t) = \sum_{i=1}^n \frac{\Delta_i \mathbf{1}_{\{T_i = t\}}}{\sum_{i=1}^n \E_i\left\{\exp \left(\bz_i^{\top}\bgamma_z +  \bbbeta_i^{\top}\bgamma_\eta \right)  \right\} \mathbf{1}_{\{T_i \geq t \}}}.
\end{eqnarray*}
The parameter vector $\bgamma = (\bgamma_z^{\top},\bgamma_{\eta}^{\top})^{\top}$ can be estimated by a one-step Newton-Raphson algorithm in an iterative manner, the $l$th iteration is
\[
\widehat{\bgamma}_l = \widehat{\bgamma}_{l - 1} + \bI_{\widehat{\bgamma}_{l - 1}}^{-1} \bS_{\widehat{\bgamma}_{l - 1}},
\]
where $\bS_{\widehat{\bgamma}_{l - 1}}$ is the score for $\bgamma$ at the $(l-1)$th iteration, and $\bI_{\widehat{\bgamma}_{l - 1}}$ is the observed information matrix. For simplicity, we omit the baseline vector $\bz_i$ from the derivation.
Following the derivations in \cite{wulfsohn1997joint}, we have
\begin{eqnarray*}
	\bS_{\bgamma} = \sum_{i=1}^{n}\left[\Delta_i  \E_i\bbbeta_i - \sum_{v=1}^{V}h_0(t_v){\E_i\left\{\bbbeta_i \exp\left( \bbbeta_i^{\top}\bgamma  \right)  \right\} \mathbf{1}_{\{T_i \geq t_v \}}} \right],
\end{eqnarray*}
where $t_v$s are the distinct observed event times, and $\bI_{\bgamma}$ can be calculated based on \cite{louis1982finding} for fast computation. Specifically,
$
\bI_{\bgamma} = \sum_{i=1}^n\bs_i(\bgamma)\bs_i(\bgamma)^{\top}/n - \bS(\bgamma)\bS(\bgamma)^{\top},
$
where $\bs_i(\bgamma)$ is the $i$th score function of $\bgamma$ obtained by taking derivatives of the log likelihood of survival (5) with respect to $\bgamma$, and $\bS(\bgamma) = \sum_{i=1}^n \bs_i(\bgamma)$.

\subsection*{Web Appendix E: Implementation Details}
\label{sec:imple}
\cite{li2010uniform} suggests $c \approx (\sum_{i=1}^n m_i)^{1/5} + 4$, which is approximately $9.13 $ for the ADNI data. We report the results based on $c=9$ since the model fittings of $c=9$ and $c=10$ are found to be close to each other.

The convergence of the EM algorithm is specified as three widely used criteria: relative and absolute changes in parameters, and relative changes in the marginal likelihood. 
The former two are used in the \texttt{R} package \texttt{JoineRML} \citep{hickey2018joinerml}, and the latter one is adopted in the \texttt{R} package \texttt{JM} \citep{rizopoulos2010jm}. 
We stop the EM algorithm when any one of the three criteria is below a predetermined threshold,
$\max\left\{\frac{\left|\widehat{\bOmega}^{(l+1)} - \widehat{\bOmega}^{(l)}\right|}{\left|\widehat{\bOmega}^{(l)}\right| + \delta_0}\right\} < \delta_1$, 
$\max\left\{\left|\widehat{\bOmega}^{(l+1)} - \widehat{\bOmega}^{(l)}\right|\right\} < \delta_2$, 
and $\frac{\left| \ell(\widehat{\bOmega}^{(l + 1)}) - \ell(\widehat{\bOmega}^{(l)})\right|}{\left| \ell(\widehat{\bOmega}^{(l)})\right| + \delta_0} < \delta_3$,
where $\widehat{\bOmega}^{(l)}$ represents the estimate of $\bOmega$ at the $\ell$th iteration, the maximum is taken over the components of $\bOmega$, $\ell(\widehat{\bOmega}^{(l)})$ is the marginal likelihood at the $\ell$th iteration, and we set $\delta_0 = 0.001$, $\delta_1 = 0.005$, $\delta_2 = 0.001$, $\delta_3 = 10^{-7}$ as suggested in \cite{booth1999maximizing}. 
Increasing the number of Monte Carlo samples may reduce the numerical errors. However, 
in practice, a large MC sample in  early iterations is not necessary as  the parameters are still far from the optima. So one may use the number of samples incrementally as the estimates move toward the truth.
In our implementation, we use Monte Carlo samples $Q = 500$ for the burn-in iterations $1 - 20$, and set $Q = 10000$ afterwards. 
We set the maximum number of EM iterations at $1000$.

\subsection*{Web Appendix F: Additional Results for Real Data Analysis}
\label{sec:supp_add_adni}

\begin{figure}[htp]
	\centering
	\scalebox{0.35}{
		\includegraphics[]{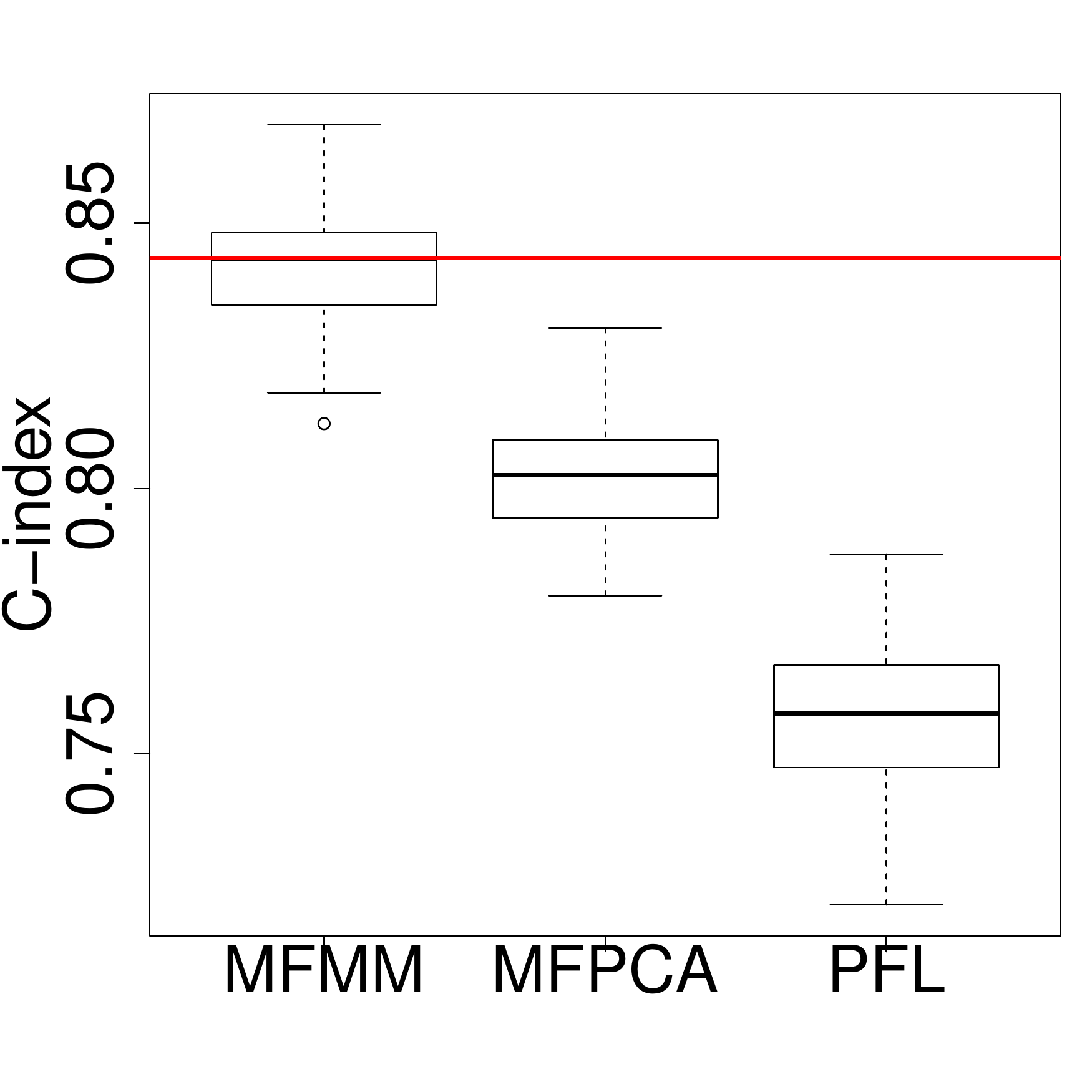}
	}
	\caption{\label{figure:c-ind} Concordance indices of different methods for predicting survival outcomes.}
\end{figure}

Figure \ref{figure:c-ind} presents the results of comparing the predictive ability of different methods in terms of the concordance index. We randomly split the ADNI data in half. The first half is used as training set, and the second half is used as testing set. 
We use PVE = 0.9 for selecting the number of principal components. The concordance indices are calculated by using \texttt{R} package \texttt{survcomp} for the testing sets. We replicate the process for 100 times.
From Figure \ref{figure:c-ind}, we can see that MFMM performs best in terms of the concordance index, followed by MFPCA; PFL is outperformed by MFPCA. The medians of the three methods are around 0.85, 0.8, and 0.75, respectively. 

\begin{figure}[htp]
	\centering
	\scalebox{0.35}{
		\includegraphics[]{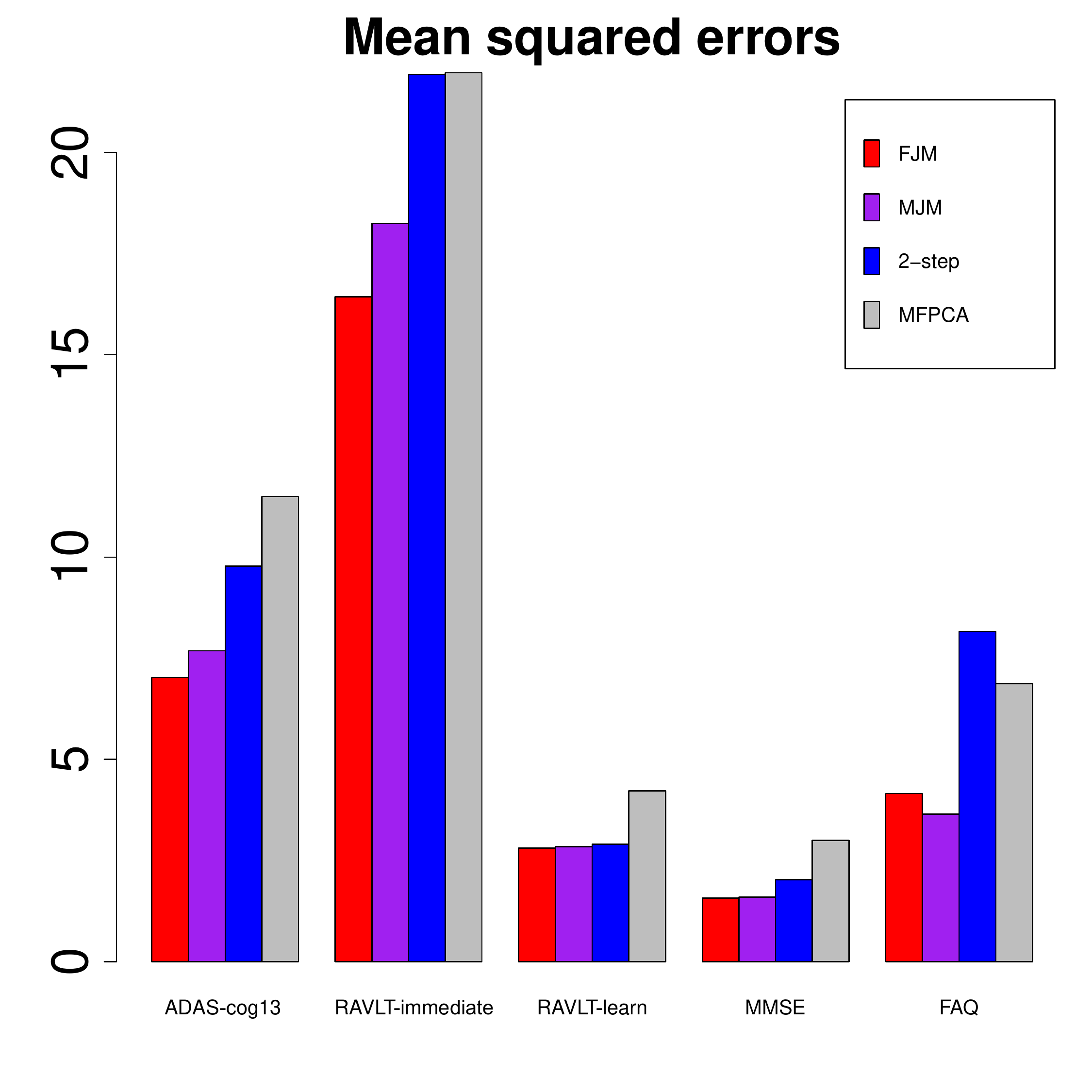}
	}
	\caption{\label{figure:mse} Mean squared errors of different methods for fitting the five longitudinal outcomes.}
\end{figure}

Figure \ref{figure:mse} visualizes the comparison of different methods for fitting the longitudinal outcomes. Here, we use mean squared errors (MSE) as the evaluation criterion, which are calculated as $\sum_{i=1}^n\sum_{k=1}^{m_i}(Y_{ijk} - \widehat{X}_{ij}(t_{ik}))^2 / \sum_{i=1}^n m_i$, where $\widehat{X}_{ij}(t_{ik})$ is the predicted latent process. We have the following findings: (1) FJM has the best overall performance due to its nonparametric flexibility, followed by MJM. Both of them outperform the two-step methods remarkably, which demonstrates the superiority of joint modeling on predicting the longitudinal outcomes. 
(2) The MFMM two-step method has smaller MSE than the MFPCA two-step method in most of the cases except for FAQ. 
This might be because MFMM effectively extracts the outcome-specific features and heterogeneous information in addition to the shared features common to all the outcomes, while MFPCA can be considered as a more nonparametric model. In summary, MFMM shows advantages over MFPCA on modeling the longitudinal data.

Table \ref{table:estimates} presents the estimates of longitudinal model components in Section 5. 
We see the advantage of using FJM compared with the two-step method (denoted as 2-step). Generally, the joint model can explain more signal variability than the two-step method. This is evidenced by the fact that estimates of eigenvalues and scaling factors are larger in magnitude than their two-step counterparts. 
On the other hand, the estimated error variances of the two-step method are larger than the ones of FJM. 
Furthermore, it is noteworthy that the scaling parameters $\beta_j$ of ADAS-Cog 13 and FAQ are $1$ and $0.67$, respectively, and those of RAVLT-immediate, RAVLT-learn and MMSE are $-1.44$, $-0.26$ and $-0.28$. These estimates are reasonable since ADAS-Cog 13 and FAQ are positively correlated in a group with higher values indicating AD progression, and the other three are positively correlated in another group with lower values suggesting of AD. The between group correlations are negative, which is reflected by the opposite signs of the scaling parameters.

\begin{figure}[htp]
	\centering
	\scalebox{0.45}{
		\includegraphics[]{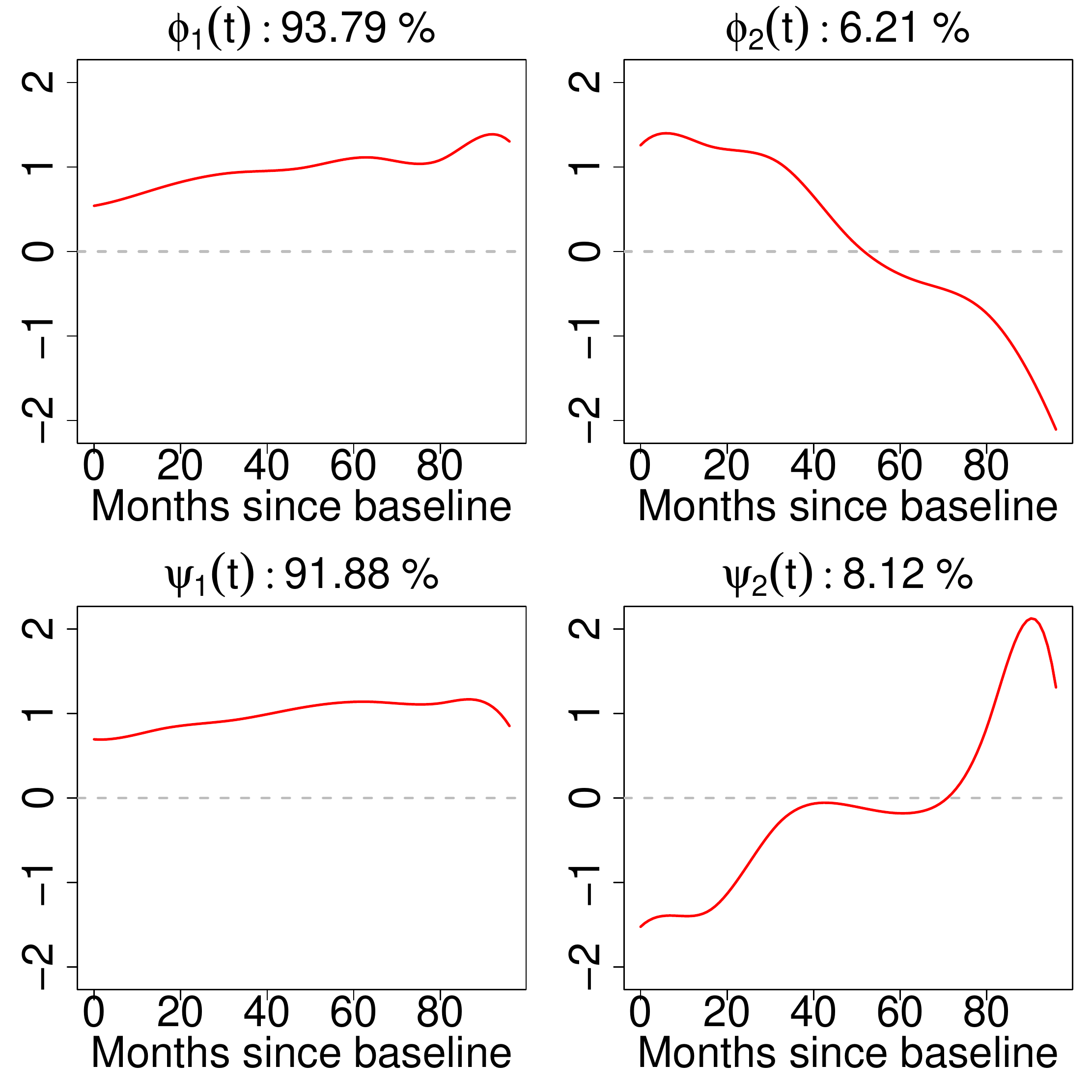}
	}
	\caption{\label{figure:eigenfunc} Estimated eigenfunctions. Percentage represents proportion of variance explained by the principal component.}
\end{figure}

Figure \ref{figure:eigenfunc} displays the estimated eigenfunctions of FJM. 
The first principal component (PC) explains approximately $94\%$ and $92\%$ of common and outcome-specific variance, respectively. The first PCs are basically vertical shifts with some fluctuations around the end of the study, which might be used as indicators of AD progression.
For example, participants who are positively loaded on them, $\xi_{i1} > 0$,  have a higher long term ADAS-Cog 13 or lower long term RAVLT-immediate than the population mean, which suggests faster AD progression.

\begin{figure}[htp]
	\centering
	\scalebox{0.35}{
		\includegraphics[]{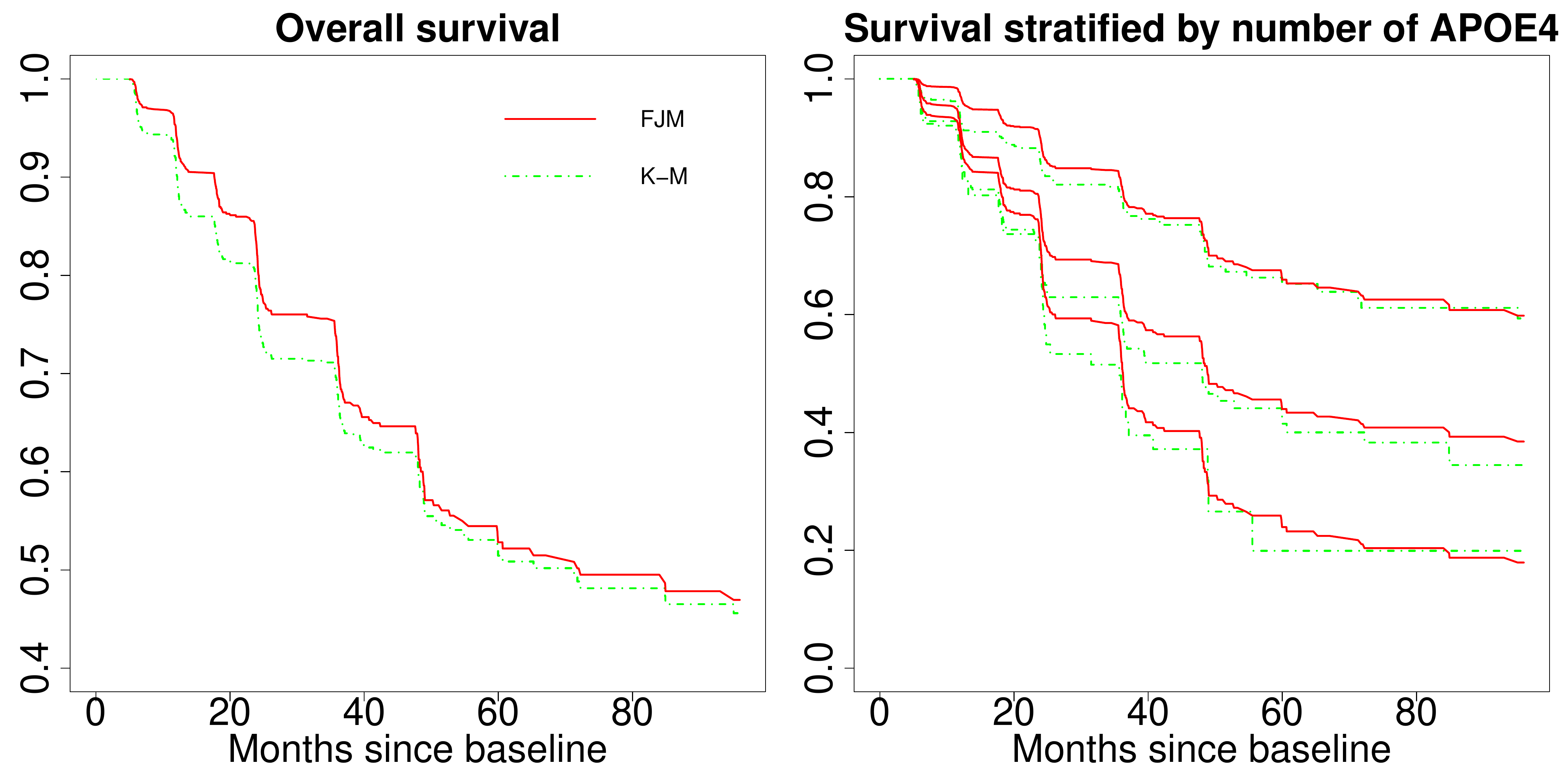}
	}
	\caption{\label{figure:survival} Survival (left panel): observed (Kaplan-Meier curve, green line) vs fitted (FJM, red line). 
		Stratified survival (right panel): observed (Kaplan-Meier curves, green lines) vs fitted (FJM, red lines).}
\end{figure}

\begin{figure}[htp]
	\centering
	\scalebox{0.35}{
		\includegraphics[]{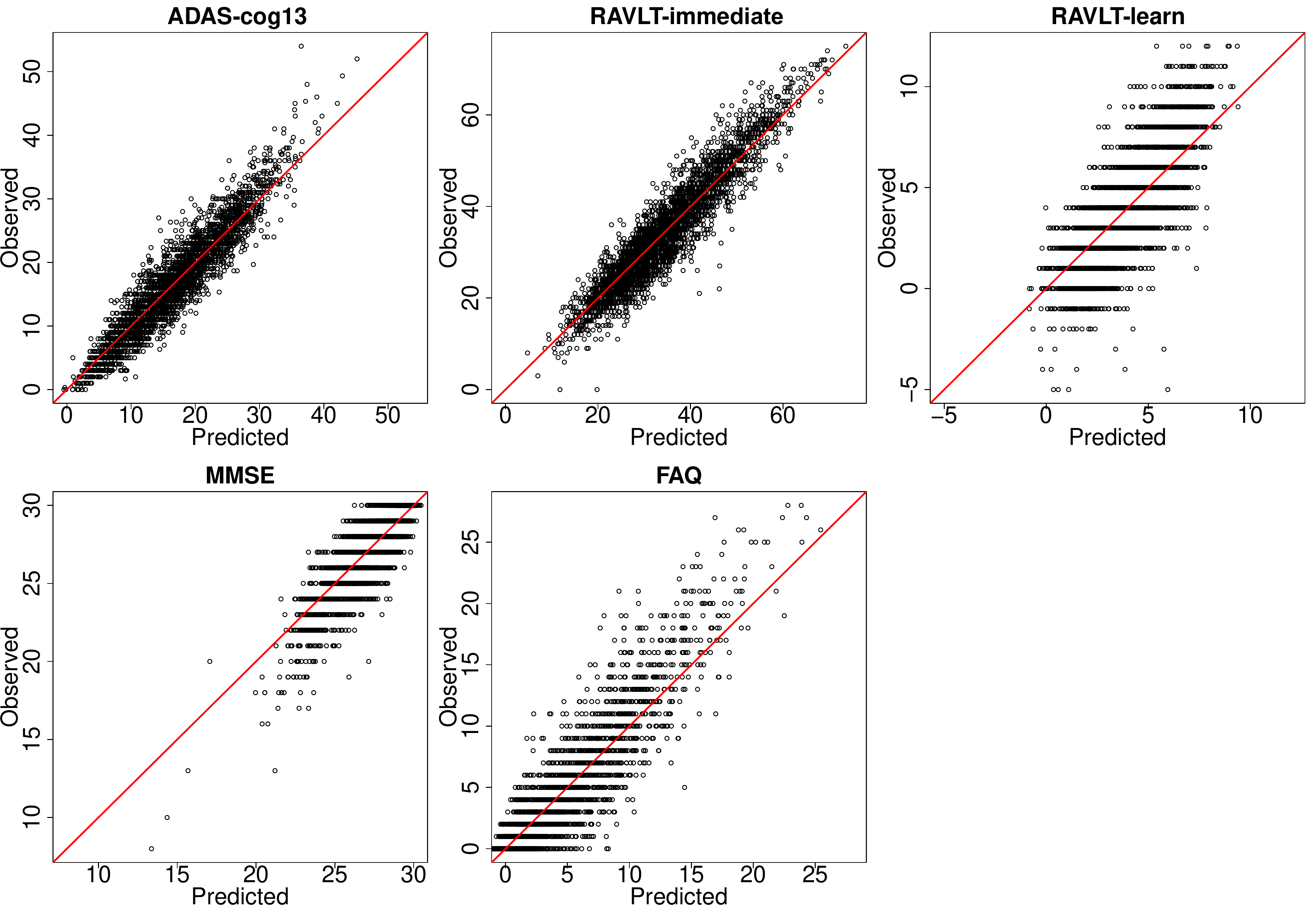}
	}
	\caption{\label{figure:long-fit} Model fitting of longitudinal outcomes: observed vs FJM fitted.}
\end{figure}

\begin{figure}[htp]
	\centering
	\scalebox{0.35}{
		\includegraphics[]{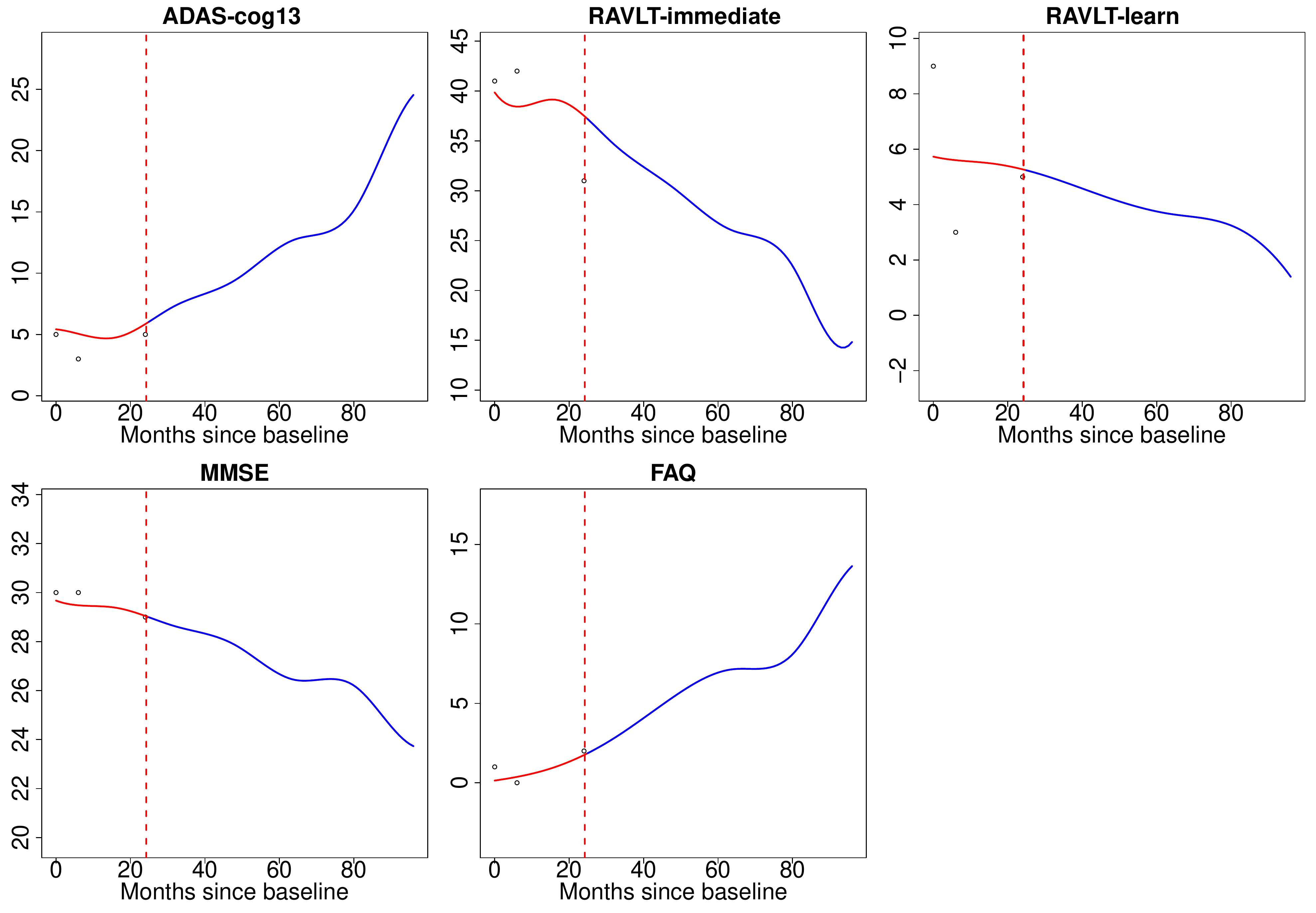}
	}
	\caption{\label{figure:long-subj} Predicted trajectories for one subject. The dashed vertical lines represent the event time. The red lines are predicted trajectories before the event time, the blue lines are predicted trajectories after the event time.}
\end{figure}

The fitted survival curve is shown in the left panel of Figure \ref{figure:survival}. 
FJM is close to the Kaplan-Meier estimate.
The right panel of Figure \ref{figure:survival} presents the survival curves of FJM and Kaplan-Meier stratified by the number of APOE4 alleles. With more APOE4 alleles, the hazard rate of AD diagnosis drastically increases as APOE4 is a well-known risk factor for AD.
Finally, Figure \ref{figure:long-fit} presents scatter plots of the observed longitudinal outcomes and the fitted ones, which show that the fitted values reasonably agree with the observed longitudinal outcomes.
In addition, Figure \ref{figure:long-subj} demonstrates the predicted latent processes of one subject.

\begin{table}[htp]
	\caption{\label{table:estimates} Estimates of other model components.}
	\centering
	\scalebox{1.00}{
		\begin{tabular}{ccc}
			\hline
			& FJM & Two-step \\
			\hline
			$d_{01}$ & 85.01 & 41.23 \\
			$d_{02}$ & 5.63 & 8.05 \\
			$d_{11}$ & 21.65 & 16.82 \\
			$d_{12}$ & 1.91 & 0.87 \\
			$\sigma^2_1$ & 9.46 & 17.76 \\
			$\sigma^2_2$ & 21.88 & 35.34 \\
			$\sigma^2_3$ & 3.32 & 3.56 \\
			$\sigma^2_4$ & 1.96 & 2.59 \\
			$\sigma^2_5$ & 5.43 & 8.12 \\
			$\beta_2$ & -1.44 & -1.21 \\
			$\beta_3$ & -0.26 & -0.25 \\
			$\beta_4$ & -0.28 & -0.21 \\
			$\beta_5$ & 0.67 & 0.33 \\
			\hline
		\end{tabular}
	}
\end{table}

\subsection*{Web Appendix G: Sensitivity Analysis for ADNI Data}
\label{sec:supp_sensitivity}
\begin{figure}[htp]
	\centering
	\scalebox{0.4}{
		\includegraphics[]{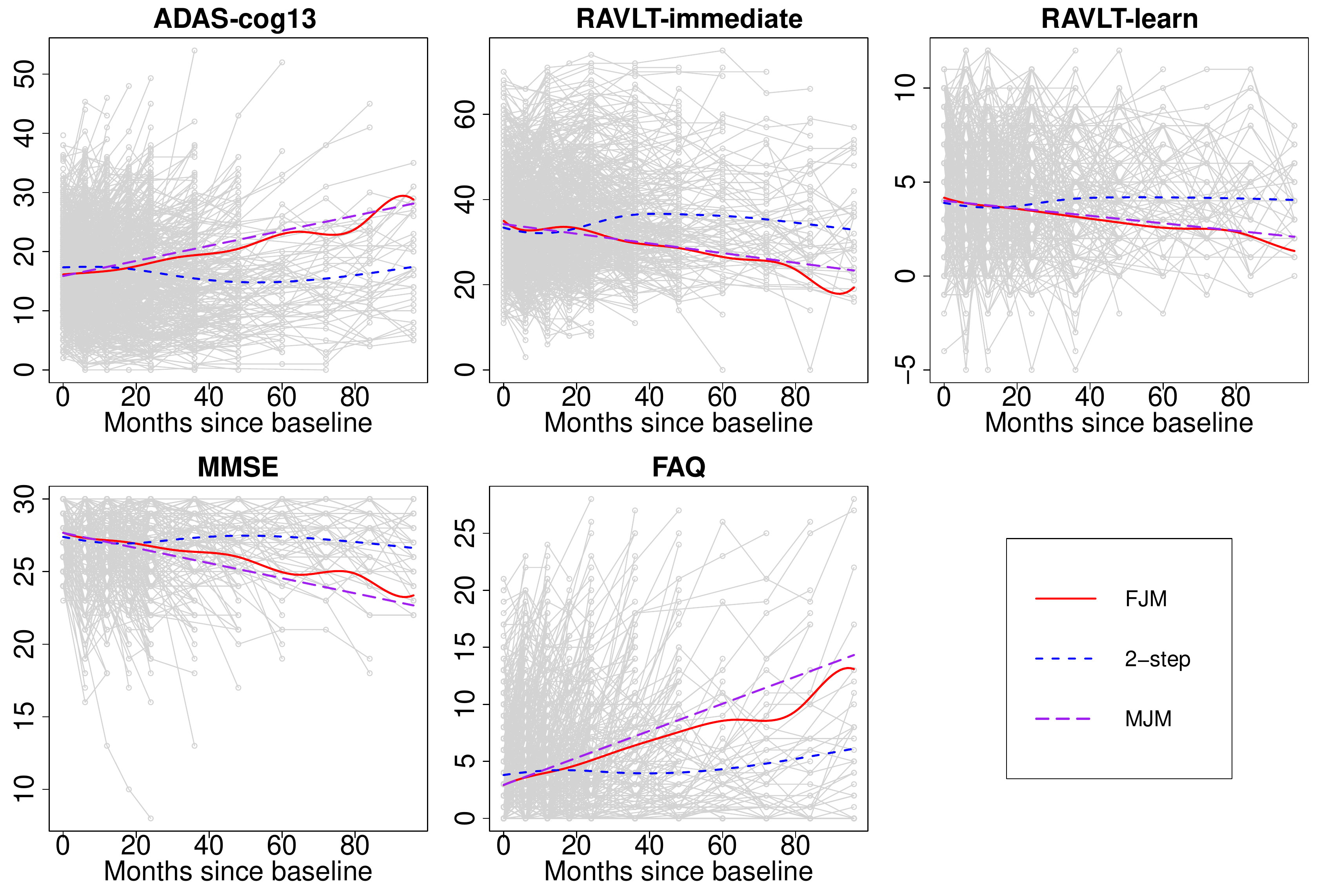}
	}
	\caption{\label{figure:mean_supp} Estimated mean functions using the ADNI data with scheduled times. Gray lines: longitudinal outcomes.}
\end{figure}

\begin{figure}[htp]
	\centering
	\scalebox{0.45}{
		\includegraphics[]{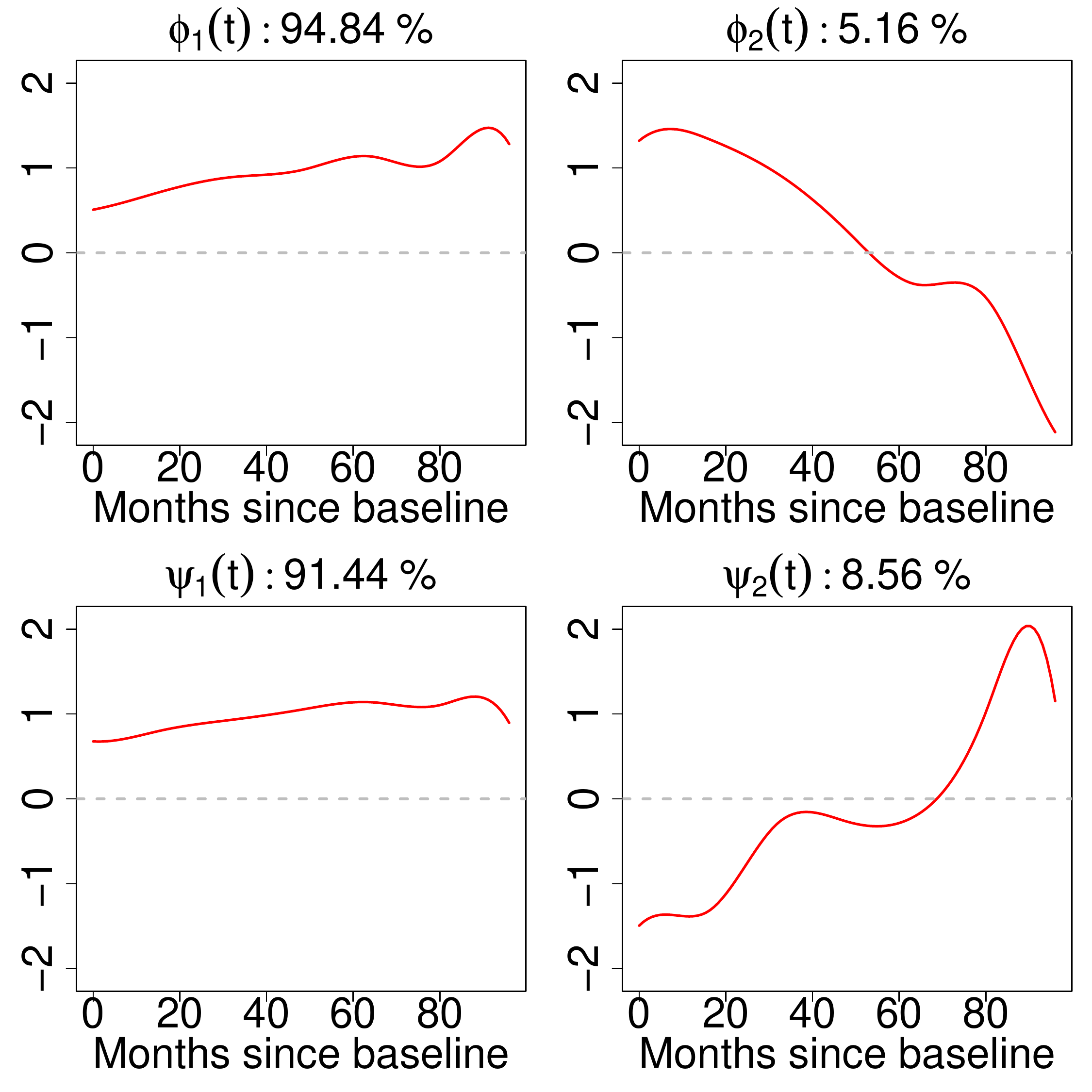}
	}
	\caption{\label{figure:eigenfunc_supp} Estimated eigenfunctions using the ADNI data with scheduled times. Percentage represents proportion of variance explained by the principal component.}
\end{figure}

We analyze the ADNI data using scheduled visit times and present the results below. The patterns of the results are almost the same as those presented in Section 5 and Web Appendix F, which shows that the results provided by the proposed FJM are robust. 
Figure \ref{figure:mean_supp} presents estimated mean functions using the ADNI data with scheduled times.
Figure \ref{figure:eigenfunc_supp} shows estimated eigenfunctions using the ADNI data with scheduled times.
Tables \ref{table:cox coeff supp} and \ref{table:estimates_supp} summarize estimates of Cox regression coefficients and other model components using the ADNI data with scheduled times, respectively.
Moreover, we shall use these model estimates as truth for generating simulation data of case 1 in Section 6.

\begin{table}[htp]
	\caption{\label{table:cox coeff supp} Estimates (standard errors) of Cox regression coefficients from functional joint model using the ADNI data with scheduled times. An asterisks indicates significance at level 0.05.}
	\centering
	\scalebox{0.9}{
		\begin{tabular}{cccc}
			\hline
			FJM & Coefficient & Estimate (standard error) & P-value \\
			\hline
			Age & $\gamma_a$ & -0.02 (0.01)$^*$ & 0.03 \\
			Gender (Female) & $\gamma_g$ & 0.27 (0.25) & 0.28  \\
			Education & $\gamma_e$ & 0.04 (0.03) & 0.29 \\
			APOE4 & $\gamma_{\epsilon}$ & 0.35 (0.17)$^*$ & 0.04 \\
			\multirow{2}{*}{Shared latent progression} &  $\gamma_{01}$ & 0.33 (0.02)$^*$ & $6e - 46$ \\
			\multirow{2}{*}{} & $\gamma_{02}$ & 0.31 (0.11)$^*$ & $0.01$ \\
			\multirow{2}{*}{ADAS-Cog 13 progression} & $\gamma_{11}$ & 0.01 (0.09) & 0.88 \\
			\multirow{2}{*}{} & $\gamma_{12}$ & -0.27 (0.23) & 0.26 \\
			\multirow{2}{*}{RAVLT-immediate progression} & $\gamma_{21}$ & 0.25 (0.10)$^*$ & 0.01 \\
			\multirow{2}{*}{} & $\gamma_{22}$ & 0.80 (0.20)$^*$ & $8e - 5$ \\
			\multirow{2}{*}{RAVLT-learn progression} & $\gamma_{31}$ & 0.05 (0.09) & 0.63 \\
			\multirow{2}{*}{} & $\gamma_{32}$ & -0.12 (0.32) & 0.72 \\
			\multirow{2}{*}{MMSE progression} & $\gamma_{41}$ & -0.04 (0.08) & 0.62 \\
			\multirow{2}{*}{} & $\gamma_{42}$ & -0.36 (0.29) & 0.21 \\
			\multirow{2}{*}{FAQ progression} & $\gamma_{51}$ & 0.12 (0.08) & 0.14 \\
			\multirow{2}{*}{} & $\gamma_{52}$ & -0.25 (0.22) & 0.25 \\
			\hline
		\end{tabular}
	}
\end{table}

\begin{table}[htp]
	\caption{\label{table:estimates_supp} Estimates of other model components using the ADNI data with scheduled times.}
	\centering
	\scalebox{1.00}{
		\begin{tabular}{ccc}
			\hline
			& FJM & Two-step \\
			\hline
			$d_{01}$ & 95.41 & 41.58 \\
			$d_{02}$ & 5.04 & 8.02 \\
			$d_{11}$ & 21.90 & 16.69 \\
			$d_{12}$ & 2.05 & 0.44 \\
			$\sigma^2_1$ & 9.49 & 17.71 \\
			$\sigma^2_2$ & 21.98 & 35.22 \\
			$\sigma^2_3$ & 3.33 & 3.55 \\
			$\sigma^2_4$ & 1.96 & 2.60 \\
			$\sigma^2_5$ & 5.49 & 8.42 \\
			$\beta_2$ & -1.44 & -1.20 \\
			$\beta_3$ & -0.26 & -0.25 \\
			$\beta_4$ & -0.28 & -0.21 \\
			$\beta_5$ & 0.67 & 0.34 \\
			\hline
		\end{tabular}
	}
\end{table}

\subsection*{Web Appendix H: Additional Results for Simulations}
\label{sec:supp_add_re}

\subsubsection*{Additional Simulation Results for Case 1}
\begin{figure}[htp]
	\centering
	\scalebox{0.55}{
		\includegraphics[]{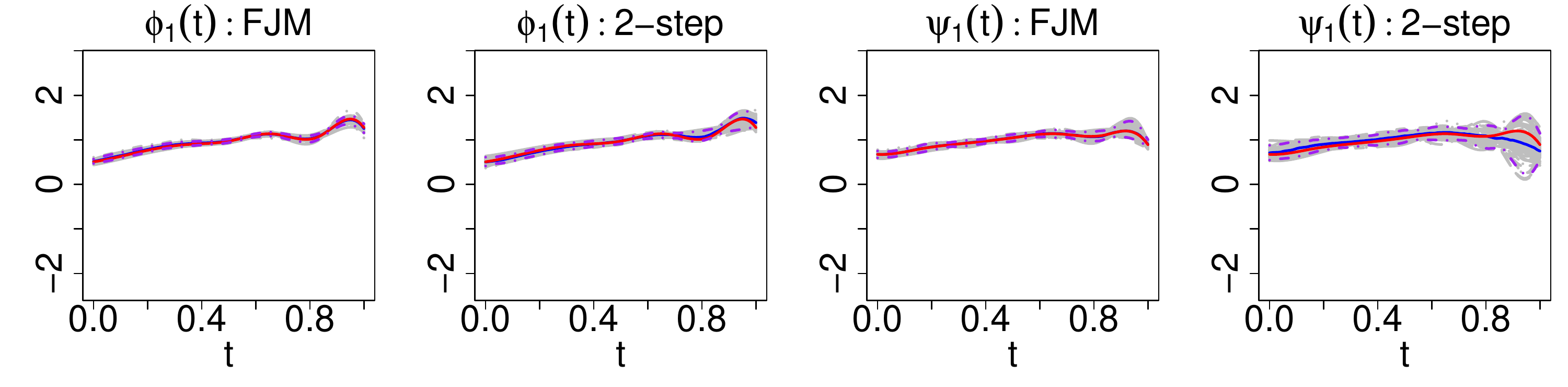}
	}
	\caption{\label{fig:eigenfunc2}Estimated eigenfunctions of $100$ replications for case 1.
		Red lines: true eigenfunctions; gray lines: estimated eigenfunctions; blue lines: medians of estimates; dashed purple lines: $95\%$ point-wise confidence bands.}
\end{figure}

\begin{figure}[htp]
	\centering
	\scalebox{0.4}{
		\includegraphics[]{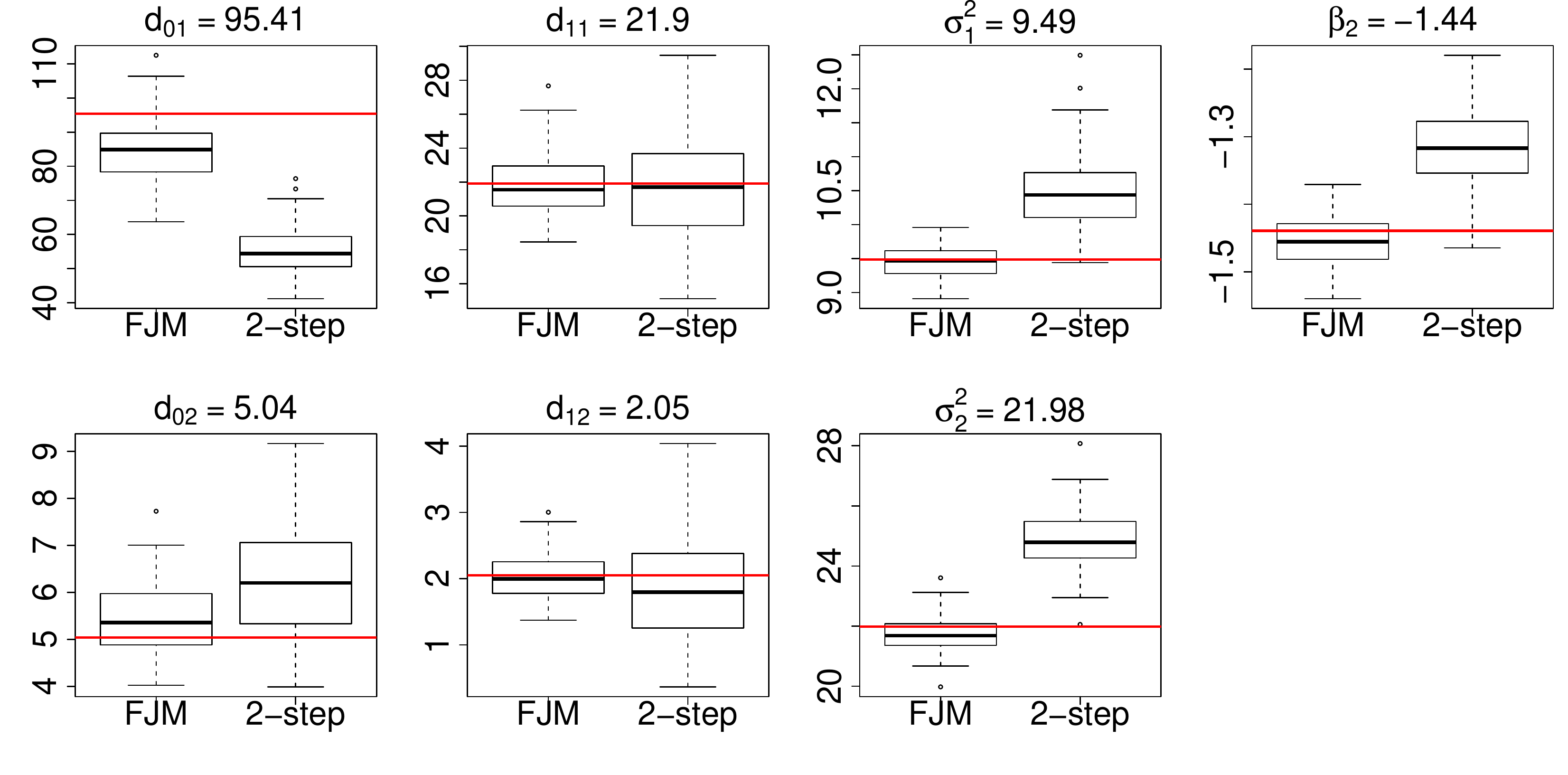}
	}
	\caption{\label{fig:boxplot2}Estimated model components of $100$ replications for case 1. The red lines represent the true parameters.}
\end{figure}

Figure \ref{fig:eigenfunc2}  presents the estimated eigenfunctions for $\phi_1(t)$ and $\psi_1(t)$.  
Figure \ref{fig:boxplot2} summarizes the estimates of model components. For various scalar parameters, FJM is reasonably close to the truth, and the two-step method shows significant bias. 
On the one hand, most of the eigenvalue estimates of the two-step method are biased toward zero. On the other hand, the  error variance estimates of the two-step method are larger than the truth, which indicates lower variance explained by the model, a phenomenon already reported in joint modeling literature.
Table \ref{table:rate1} presents the results of rank selection.
The cases of FJM with misspecified ranks using AIC tend to select $3$ principal components while the truth is $2$. 

\begin{table}[htp]
	\caption{\label{table:rate1} Proportions of rank selections among $100$ replications for case 1. $L_j$ is the true rank, and $\widehat{L}_j$ is the selected rank, $j = 0, 1$.}
	\centering
	\scalebox{1.0}{
		\begin{tabular}{ccccccccc}
			\hline
			\multirow{3}{*}{} & \multicolumn{4}{c}{AIC} & \multicolumn{4}{c}{BIC} \\
			\multirow{3}{*}{} & \multicolumn{2}{c}{$L_0$} & \multicolumn{2}{c}{$L_1$} & \multicolumn{2}{c}{$L_0$} & \multicolumn{2}{c}{$L_1$} \\
			\multirow{3}{*}{} & 2-step & FJM & 2-step & FJM & 2-step & FJM & 2-step & FJM \\
			\hline
			$\widehat{L}_j < L_j$   & 0.00 & 0.00 & 0.01 & 0.00 & 0.00 & 0.00 & 0.12 & 0.00 \\
			$\widehat{L}_j = L_j$   & 0.39 & 0.92 & 0.34 & 0.95 & 0.40 & 1.00 & 0.38 & 1.00 \\
			$\widehat{L}_j > L_j$   & 0.61 & 0.08 & 0.65 & 0.05 & 0.60 & 0.00 & 0.50 & 0.00 \\
			\hline
		\end{tabular}
	}
\end{table}

\subsubsection*{Simulation Settings for Case 2}
In this scenario, we use an alternative setting of model components to further investigate the numerical property of the proposed methods. 
Again, we specify the number of longitudinal outcomes as $J = 2$ and simulate the longitudinal data as before.
In particular, we let $\mu_1(t) = 5 \sin (2 \pi t)$ and $\mu_2(t) = 5 \cos (2 \pi t)$ be the two mean functions.
We set two principal components $L_0 = L_1 = 2$ for the two covariances.
We specify eigenfunctions of $\C_0(s, t)$ as $\phi_1(t) = \sqrt{2} \sin (\pi t)$, $\phi_2(t) = \sqrt{2} \cos (-3 \pi t)$, and the eigen scores $\xi_{i \ell}$'s are generated as before with $d_{0 \ell} = 1 / 2^{\ell - 1}$ for $\ell = 1, 2$.
We use $\psi_1(t) = \sqrt{2} \cos (\pi t)$, $\psi_2(t) = \sqrt{2} \cos (2 \pi t)$ as the two eigenfunctions of $\C_1(s, t)$, and the outcome-specific eigen scores $\zeta_{i j \ell}$'s are generated with $d_{1 \ell} = d_{0 \ell} / 2$ for $\ell = 1, 2$. 
The scaling parameters are set as $\beta_1 = 1$ and $\beta_2 = -1$. 
The white noise $\epsilon_{ijk}$s are normals with zero mean and error variance $\sigma_j^2$. 
We specify signal-to-noise ratio (SNR) as $1.5$, i.e., $\sigma_j^2 = (\sum_{\ell=1}^2 d_{0 \ell} + \sum_{\ell=1}^2 d_{1 \ell}) / 3$ for $j = 1, 2$. The observed time points $t_{ijk} = t_{ik}$ are $11$ equally-spaced points in the interval $[0,1]$, which mimics the ADNI study. 

The time-to-event data are generated as in case 1 of Section 6, but the Cox coefficients are $\bgamma_0 = (1, 0.5)^{\top}$, $\bgamma_{11} = \bgamma_{12} = (0.2, 0.1)^{\top}$. The censoring rate is around $30\%$. 
We generate data with $800$ subjects, and the average number of observations per subject is around 5, which are close to the ADNI study.

\subsubsection*{Simulation Results for Case 2}
\begin{figure}[htp]
	\centering
	\scalebox{0.27}{
		\includegraphics[]{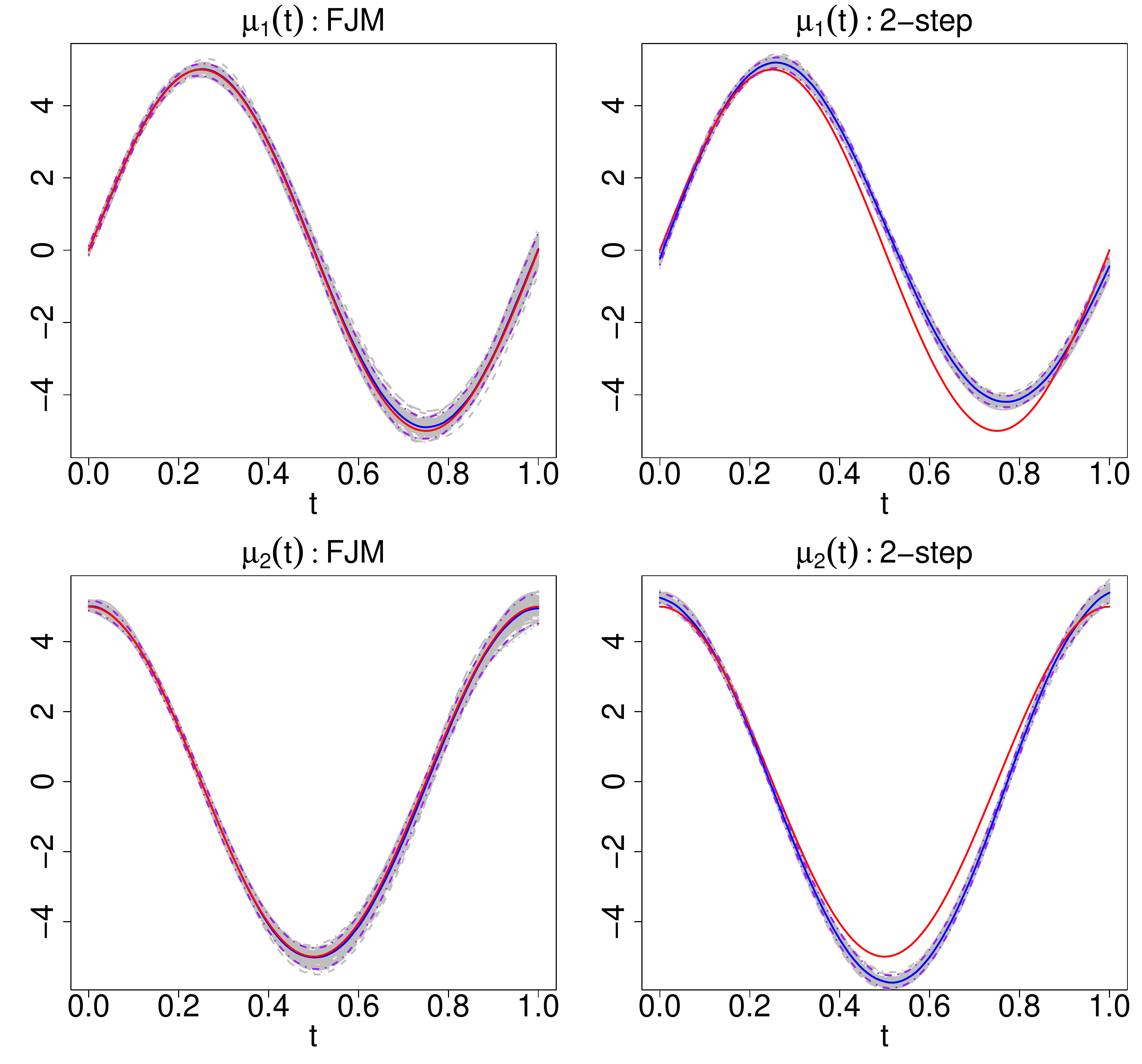}
	}
	\caption{\label{fig:mean1}Estimated mean functions of $100$ replications for case 2. The gray lines are the estimates. The red lines are the true means. The blue lines represent the medians of the estimates. The purple lines are corresponding $95\%$ point-wise confidence bands.}
\end{figure}

Figure \ref{fig:mean1} presents the estimated mean functions for case 2. 
The medians of FJM are almost identical to the truth and its $95\%$ point-wise confidence bands succeed in covering the true means. By contrast, the two-step method has obvious bias when the subjects have fewer observations along the way, and its $95\%$ confidence bands cannot always cover the truth.

\begin{figure}[htp]
	\centering
	\scalebox{0.5}{
		\includegraphics[]{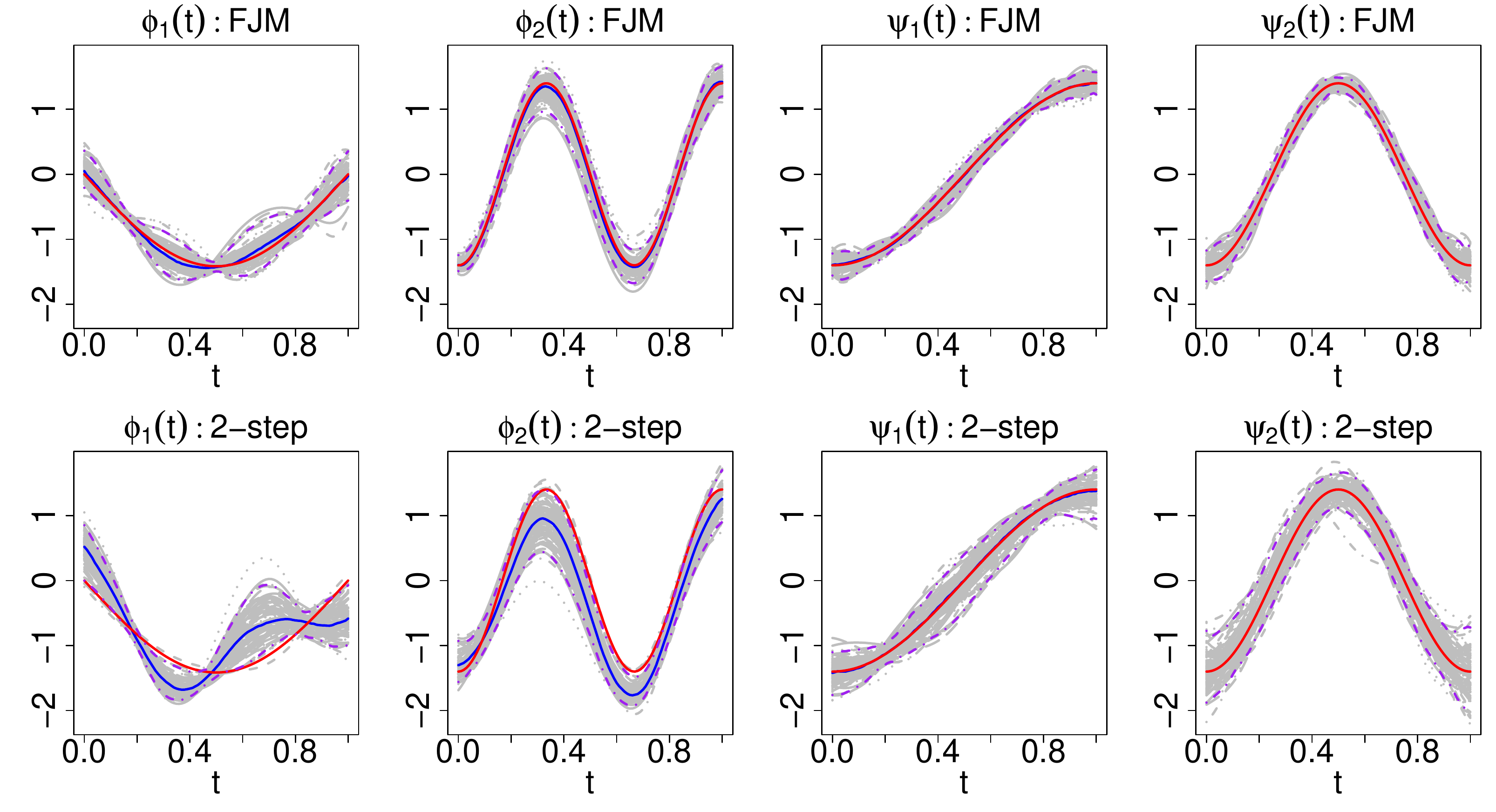}
	}
	\caption{\label{fig:eigenfunc1}Estimated eigenfunctions of $100$ replications for case 2. The gray lines are the estimates. The red lines are the true eigenfunctions. The blue lines represent the medians of the estimates. The purple lines are corresponding $95\%$ point-wise confidence bands.}
\end{figure}

Figure \ref{fig:eigenfunc1} shows the estimated eigenfunctions for case 2. 
For $\phi_1(t)$, the median of FJM is close to the truth and the true $\phi_1(t)$ lies within the $95\%$ point-wise confidence band. By contrast, the two-step estimates are far from the truth, and most of them even show a different shape. As a result, the median and confidence band of the two-step method indicate large bias. 
For $\phi_2(t)$, while FJM still provides reasonable estimates, the two-step method again yields biased estimates. It seems that the two-step method can resemble the shape of $\phi_2(t)$ but with inaccurate magnitude. Therefore, the median shows systematic bias, and the truth lies in the margin area of the $95\%$ confidence band. 
For $\psi_1(t)$ and $\psi_2(t)$, both FJM and the two-step method estimate the eigenfunctions well, but FJM gives much narrower confidence bands.

\begin{figure}[htp]
	\centering
	\scalebox{0.5}{
		\includegraphics[]{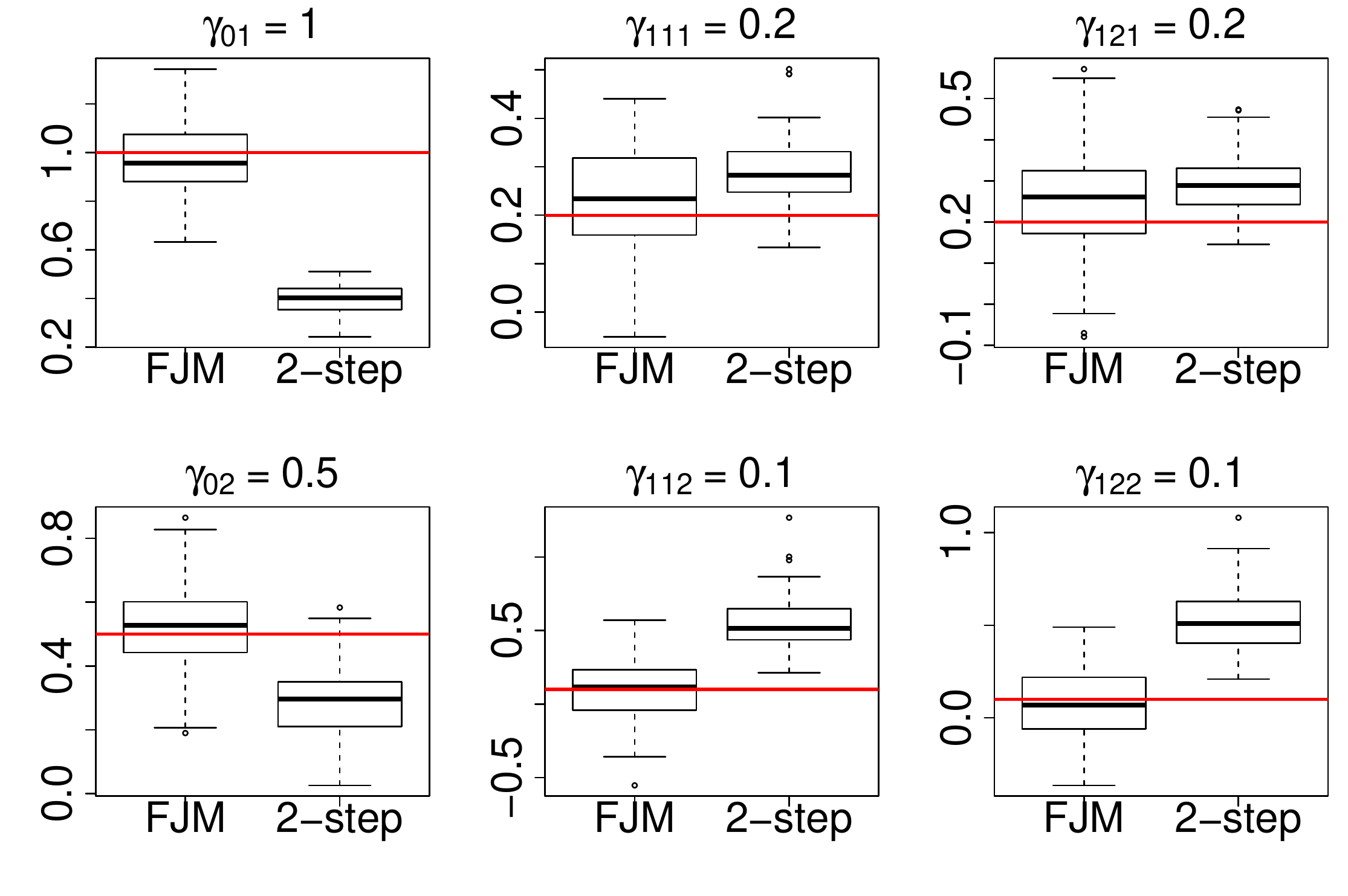}
	}
	\caption{\label{fig:boxplot_cox1}Estimated Cox coefficients of $100$ replications for case 2. The red lines represent the true parameters.}
\end{figure}

\begin{figure}[htp]
	\centering
	\scalebox{0.45}{
		\includegraphics[]{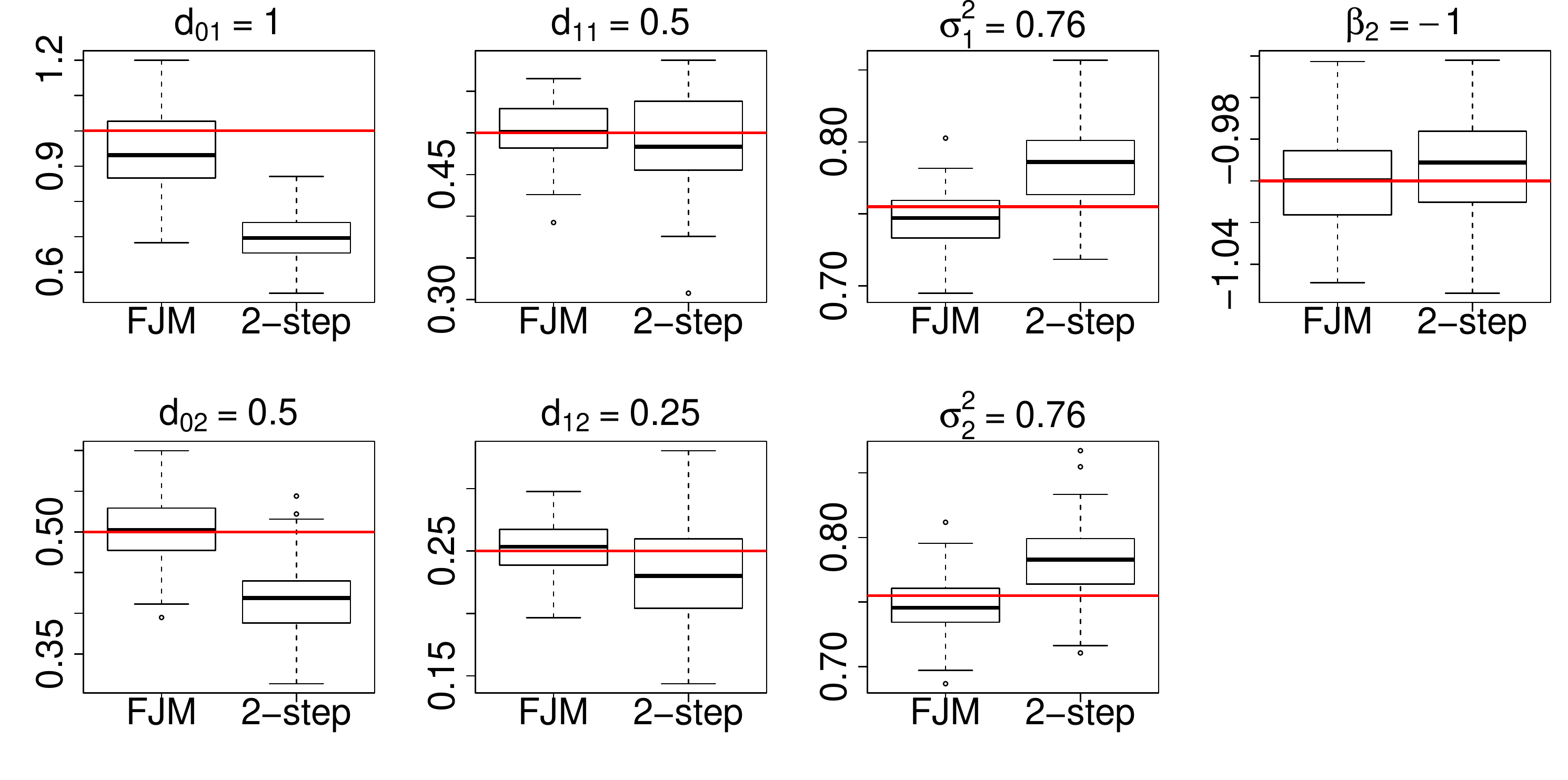}
	}
	\caption{\label{fig:boxplot1}Estimated model components of $100$ replications for case 2. The red lines represent the true parameters.}
\end{figure}

Figures \ref{fig:boxplot_cox1} and \ref{fig:boxplot1} present the estimated scalar parameters for case 2. The model estimates of FJM are very close to the truth, and much more accurate than the two-step method as well.
In particular, the boxplots of $\gamma_{121}$, $\gamma_{122}$ and $\sigma_2^2$ show similar trend as $\gamma_{111}$, $\gamma_{112}$ and $\sigma_1^2$ even though they are estimated separately. 

Table \ref{table:rate2} summarizes the results of rank selections for case 2. The pattern is consistent with case 1 of Section 6:  both AIC and BIC for FJM achieve high rates of correctly selecting the number of principal components.

\begin{table}[htp]
	\caption{\label{table:rate2} Proportions of rank selections among $100$ replications for case 2. $L_j$ is the true rank, and $\widehat{L}_j$ is the selected rank, $j = 0, 1$.}
	\centering
	\scalebox{1.0}{
		\begin{tabular}{ccccccccc}
			\hline
			\multirow{3}{*}{} & \multicolumn{4}{c}{AIC} & \multicolumn{4}{c}{BIC} \\
			\multirow{3}{*}{} & \multicolumn{2}{c}{$L_0$} & \multicolumn{2}{c}{$L_1$} & \multicolumn{2}{c}{$L_0$} & \multicolumn{2}{c}{$L_1$} \\
			\multirow{3}{*}{} & 2-step & FJM & 2-step & FJM & 2-step & FJM & 2-step & FJM \\
			\hline
			$\widehat{L}_j < L_j$   & 0.12 & 0.00 & 0.12 & 0.00 & 0.20 & 0.00 & 0.47 & 0.00 \\
			$\widehat{L}_j = L_j$   & 0.14 & 0.94 & 0.18 & 0.91 & 0.15 & 1.00 & 0.26 & 1.00 \\
			$\widehat{L}_j > L_j$   & 0.74 & 0.06 & 0.70 & 0.00 & 0.65 & 0.00 & 0.27 & 0.00 \\
			\hline
		\end{tabular}
	}
\end{table}

\newpage
\bibliography{ref}

\begin{thebibliography}{}

\bibitem[\protect\citeauthoryear{{Alzheimer's Association}}{{Alzheimer's
  Association}}{2019}]{alzheimer20192019}
{Alzheimer's Association} (2019).
\newblock 2019 {Alzheimer's} disease facts and figures.
\newblock {\em Alzheimer's \& Dementia\/}~{\em 15\/}(3), 321--387.

\bibitem[\protect\citeauthoryear{Booth and Hobert}{Booth and
  Hobert}{1999}]{booth1999maximizing}
Booth, J.~G. and J.~P. Hobert (1999).
\newblock Maximizing generalized linear mixed model likelihoods with an
  automated monte carlo {EM} algorithm.
\newblock {\em Journal of the Royal Statistical Society: Series B (Statistical
  Methodology)\/}~{\em 61\/}(1), 265--285.

\bibitem[\protect\citeauthoryear{de~Boor}{de~Boor}{1978}]{deBoor:78}
de~Boor, C. (1978).
\newblock {\em A Practical Guide to Splines}.
\newblock Berlin: Springer.

\bibitem[\protect\citeauthoryear{De~Gruttola and Tu}{De~Gruttola and
  Tu}{1994}]{de1994modelling}
De~Gruttola, V. and X.~M. Tu (1994).
\newblock Modelling progression of {CD}4-lymphocyte count and its relationship
  to survival time.
\newblock {\em Biometrics\/}, 1003--1014.

\bibitem[\protect\citeauthoryear{Di, Crainiceanu, Caffo, and Punjabi}{Di
  et~al.}{2009}]{di2009multilevel}
Di, C.-Z., C.~M. Crainiceanu, B.~S. Caffo, and N.~M. Punjabi (2009).
\newblock Multilevel functional principal component analysis.
\newblock {\em The Annals of Applied Statistics\/}~{\em 3\/}(1), 458--488.

\bibitem[\protect\citeauthoryear{Eilers and Marx}{Eilers and
  Marx}{1996}]{Eilers:96}
Eilers, P. and B.~Marx (1996).
\newblock {Flexible smoothing with B-splines and penalties (with Discussion)}.
\newblock {\em Statistical Science\/}~{\em 11\/}(2), 89--121.

\bibitem[\protect\citeauthoryear{Fleisher, Sowell, Taylor, Gamst, Petersen,
  Thal, et~al.}{Fleisher et~al.}{2007}]{fleisher2007clinical}
Fleisher, A., B.~Sowell, C.~Taylor, A.~Gamst, R.~C. Petersen, L.~Thal, et~al.
  (2007).
\newblock Clinical predictors of progression to {A}lzheimer disease in amnestic
  mild cognitive impairment.
\newblock {\em Neurology\/}~{\em 68\/}(19), 1588--1595.

\bibitem[\protect\citeauthoryear{Happ and Greven}{Happ and
  Greven}{2018}]{happ2018multivariate}
Happ, C. and S.~Greven (2018).
\newblock Multivariate functional principal component analysis for data
  observed on different (dimensional) domains.
\newblock {\em Journal of the American Statistical Association\/}~{\em
  113\/}(522), 649--659.

\bibitem[\protect\citeauthoryear{Harrell~Jr}{Harrell~Jr}{2015}]{harrell2015regression}
Harrell~Jr, F.~E. (2015).
\newblock {\em Regression modeling strategies: with applications to linear
  models, logistic and ordinal regression, and survival analysis}.
\newblock Springer.

\bibitem[\protect\citeauthoryear{Henderson, Diggle, and Dobson}{Henderson
  et~al.}{2000}]{henderson2000joint}
Henderson, R., P.~Diggle, and A.~Dobson (2000).
\newblock Joint modelling of longitudinal measurements and event time data.
\newblock {\em Biostatistics\/}~{\em 1\/}(4), 465--480.

\bibitem[\protect\citeauthoryear{Hickey, Philipson, Jorgensen, and
  Kolamunnage-Dona}{Hickey et~al.}{2018}]{hickey2018joinerml}
Hickey, G.~L., P.~Philipson, A.~Jorgensen, and R.~Kolamunnage-Dona (2018).
\newblock joine{RML}: a joint model and software package for time-to-event and
  multivariate longitudinal outcomes.
\newblock {\em BMC Medical Research Methodology\/}~{\em 18\/}(1), 50.

\bibitem[\protect\citeauthoryear{Huang, Li, and Guan}{Huang
  et~al.}{2014}]{huang2014joint}
Huang, H., Y.~Li, and Y.~Guan (2014).
\newblock Joint modeling and clustering paired generalized longitudinal
  trajectories with application to cocaine abuse treatment data.
\newblock {\em Journal of the American Statistical Association\/}~{\em
  109\/}(508), 1412--1424.

\bibitem[\protect\citeauthoryear{Kong, Giovanello, Wang, Lin, Lee, Fan,
  et~al.}{Kong et~al.}{2015}]{kong2015predicting}
Kong, D., K.~S. Giovanello, Y.~Wang, W.~Lin, E.~Lee, Y.~Fan, et~al. (2015).
\newblock Predicting alzheimer's disease using combined imaging-whole genome
  snp data.
\newblock {\em Journal of Alzheimer's Disease\/}~{\em 46\/}(3), 695--702.

\bibitem[\protect\citeauthoryear{Kong, Ibrahim, Lee, and Zhu}{Kong
  et~al.}{2018}]{kong2018flcrm}
Kong, D., J.~G. Ibrahim, E.~Lee, and H.~Zhu (2018).
\newblock {FLCRM}: Functional linear {C}ox regression model.
\newblock {\em Biometrics\/}~{\em 74\/}(1), 109--117.

\bibitem[\protect\citeauthoryear{Kong, Xue, Yao, and Zhang}{Kong
  et~al.}{2016}]{kong2016partially}
Kong, D., K.~Xue, F.~Yao, and H.~H. Zhang (2016).
\newblock Partially functional linear regression in high dimensions.
\newblock {\em Biometrika\/}~{\em 103\/}(1), 147--159.

\bibitem[\protect\citeauthoryear{Li, Xiao, and Luo}{Li
  et~al.}{2020}]{li2018fast}
Li, C., L.~Xiao, and S.~Luo (2020).
\newblock Fast covariance estimation for multivariate sparse functional data.
\newblock {\em Stat\/}~{\em 9\/}(1), e245.

\bibitem[\protect\citeauthoryear{Li, Chan, Doody, Quinn, and Luo}{Li
  et~al.}{2017}]{li2017prediction}
Li, K., W.~Chan, R.~S. Doody, J.~Quinn, and S.~Luo (2017).
\newblock Prediction of conversion to {A}lzheimer's disease with longitudinal
  measures and time-to-event data.
\newblock {\em Journal of Alzheimer's Disease\/}~{\em 58\/}(2), 361--371.

\bibitem[\protect\citeauthoryear{Li, Wu, and Sun}{Li
  et~al.}{2019}]{li2019penalized}
Li, S., Q.~Wu, and J.~Sun (2019).
\newblock Penalized estimation of semiparametric transformation models with
  interval-censored data and application to {A}lzheimer's disease.
\newblock {\em Statistical Methods in Medical Research\/}, 0962280219884720.

\bibitem[\protect\citeauthoryear{Li and Hsing}{Li and
  Hsing}{2010}]{li2010uniform}
Li, Y. and T.~Hsing (2010).
\newblock Uniform convergence rates for nonparametric regression and principal
  component analysis in functional/longitudinal data.
\newblock {\em The Annals of Statistics\/}~{\em 38\/}(6), 3321--3351.

\bibitem[\protect\citeauthoryear{Lin, McCulloch, and Mayne}{Lin
  et~al.}{2002}]{lin2002maximum}
Lin, H., C.~E. McCulloch, and S.~T. Mayne (2002).
\newblock Maximum likelihood estimation in the joint analysis of time-to-event
  and multiple longitudinal variables.
\newblock {\em Statistics in Medicine\/}~{\em 21\/}(16), 2369--2382.

\bibitem[\protect\citeauthoryear{Louis}{Louis}{1982}]{louis1982finding}
Louis, T.~A. (1982).
\newblock Finding the observed information matrix when using the {EM}
  algorithm.
\newblock {\em Journal of the Royal Statistical Society: Series B
  (Methodological)\/}~{\em 44\/}(2), 226--233.

\bibitem[\protect\citeauthoryear{Rizopoulos}{Rizopoulos}{2010}]{rizopoulos2010jm}
Rizopoulos, D.~D. (2010).
\newblock {JM}: An {R} package for the joint modelling of longitudinal and
  time-to-event data.
\newblock {\em Journal of Statistical Software (Online)\/}~{\em 35\/}(9),
  1--33.

\bibitem[\protect\citeauthoryear{Tsiatis and Davidian}{Tsiatis and
  Davidian}{2004}]{tsiatis2004joint}
Tsiatis, A.~A. and M.~Davidian (2004).
\newblock Joint modeling of longitudinal and time-to-event data: An overview.
\newblock {\em Statistica Sinica\/}, 809--834.

\bibitem[\protect\citeauthoryear{Verbeke, Fieuws, Molenberghs, and
  Davidian}{Verbeke et~al.}{2014}]{Verbeke2014}
Verbeke, G., S.~Fieuws, G.~Molenberghs, and M.~Davidian (2014).
\newblock The analysis of multivariate longitudinal data: A review.
\newblock {\em Statistical Methods in Medical Research\/}~{\em 23\/}(1),
  42--59.

\bibitem[\protect\citeauthoryear{Weiner, Veitch, Aisen, Beckett, Cairns, Green,
  et~al.}{Weiner et~al.}{2017}]{weiner2017recent}
Weiner, M.~W., D.~P. Veitch, P.~S. Aisen, L.~A. Beckett, N.~J. Cairns, R.~C.
  Green, et~al. (2017).
\newblock Recent publications from the {A}lzheimer's disease neuroimaging
  initiative: Reviewing progress toward improved ad clinical trials.
\newblock {\em Alzheimer's \& {D}ementia\/}~{\em 13\/}(4), e1--e85.

\bibitem[\protect\citeauthoryear{Wu and Carroll}{Wu and
  Carroll}{1988}]{wu1988estimation}
Wu, M.~C. and R.~J. Carroll (1988).
\newblock Estimation and comparison of changes in the presence of informative
  right censoring by modeling the censoring process.
\newblock {\em Biometrics\/}, 175--188.

\bibitem[\protect\citeauthoryear{Wulfsohn and Tsiatis}{Wulfsohn and
  Tsiatis}{1997}]{wulfsohn1997joint}
Wulfsohn, M.~S. and A.~A. Tsiatis (1997).
\newblock A joint model for survival and longitudinal data measured with error.
\newblock {\em Biometrics\/}, 330--339.

\bibitem[\protect\citeauthoryear{Yan, Lin, and Huang}{Yan
  et~al.}{2017}]{yan2017dynamic}
Yan, F., X.~Lin, and X.~Huang (2017).
\newblock Dynamic prediction of disease progression for leukemia patients by
  functional principal component analysis of longitudinal expression levels of
  an oncogene.
\newblock {\em The Annals of Applied Statistics\/}~{\em 11\/}(3), 1649--1670.

\bibitem[\protect\citeauthoryear{Yao}{Yao}{2007}]{yao2007functional}
Yao, F. (2007).
\newblock Functional principal component analysis for longitudinal and survival
  data.
\newblock {\em Statistica Sinica\/}, 965--983.

\bibitem[\protect\citeauthoryear{Yao, M\"uller, and Wang}{Yao
  et~al.}{2005}]{Yao:05}
Yao, F., H.-G. M\"uller, and J.-L. Wang (2005).
\newblock Functional data analysis for sparse longitudinal data.
\newblock {\em Journal of the American Statistical Association\/}~{\em
  100\/}(470), 577--590.

\bibitem[\protect\citeauthoryear{Ye, Li, and Guan}{Ye
  et~al.}{2015}]{ye2015joint}
Ye, J., Y.~Li, and Y.~Guan (2015).
\newblock Joint modeling of longitudinal drug using pattern and time to first
  relapse in cocaine dependence treatment data.
\newblock {\em The Annals of Applied Statistics\/}~{\em 9\/}(3), 1621--1642.

\bibitem[\protect\citeauthoryear{Zhou, Shen, and Wolfe}{Zhou
  et~al.}{1998}]{zhou1998local}
Zhou, S., X.~Shen, and D.~Wolfe (1998).
\newblock Local asymptotics for regression splines and confidence regions.
\newblock {\em The Annals of Statistics\/}~{\em 26\/}(5), 1760--1782.

\end{thebibliography}
\bibliographystyle{chicago}
\end{document}